\documentclass[12pt,preprint]{aastex}
\usepackage{psfig}
\begin{document}
\title{Systematic effects and a new determination of the primordial 
abundance of $^4$He and dY/dZ from observations of blue compact galaxies}
\author{Yuri I. Izotov\footnote{Visiting astronomer, 
Kitt Peak National Observatory, National Optical Astronomical Observatory,
operated by the Association of Universities for Research in Astronomy,
Inc., under contract with the National Science Foundation.}}
\affil{Main Astronomical Observatory, Ukrainian National Academy of Sciences,
27 Zabolotnoho str., Kyiv 03680, Ukraine}
\email{izotov@mao.kiev.ua}
\and
\author{Trinh X. Thuan\footnotemark[1]}
\affil{Astronomy Department, University of Virginia, Charlottesville,
VA 22903}
\email{txt@virginia.edu}

\begin{abstract}
    We use spectroscopic observations of a sample of 82 
H {\sc ii} regions in 76 blue compact galaxies to determine the 
primordial helium abundance $Y_p$ and the slope $dY/dZ$ from the $Y$ -- O/H
linear regression. To improve the accuracy of the $dY/dZ$ measurement,  
we have included new spectrophotometric observations 
of 33 H {\sc ii} regions which span a large metallicity range, 
with oxygen abundance 12 + log(O/H) varying between 7.43 
and 8.30 ($Z_\odot$/30 $\leq$$Z$$\leq$ $Z_\odot$/4). Most of the 
new galaxies were 
selected from the First Byurakan, the Hamburg/SAO and the University of 
Michigan objective prism surveys. 
For a subsample of 7 H {\sc ii} regions, we derive the He mass fraction
taking into account known systematic effects, including collisional and
fluorescent enhancements of He {\sc i} emission lines, collisional 
excitation of hydrogen emission, underlying stellar He {\sc i} 
absorption and the difference between the temperatures $T_e$(He {\sc ii})
in the He$^+$ zone and $T_e$(O {\sc iii}) derived from the 
collisionally excited [O {\sc iii}] lines. We find that the net result 
of all the systematic effects combined is small, changing the He mass 
fraction by less than 0.6\%.
     By extrapolating the $Y$ vs. O/H linear regression to
O/H = 0 for 7 H {\sc ii} regions of this subsample, we obtain 
$Y_p$ = 0.2421$\pm$0.0021 and $dY/d$O = 5.7$\pm$1.8, which corresponds to
$dY/dZ$ = 3.7$\pm$1.2, assuming the oxygen mass fraction to be 
O=0.66$Z$. In the framework
of the standard Big Bang nucleosynthesis theory, this $Y_p$ 
corresponds to $\Omega_b$$h^2$ = 0.012$^{+0.003}_{-0.002}$, where 
$h$ is the Hubble constant in units of 100 km s$^{-1}$ Mpc$^{-1}$. This   
is smaller at the 2$\sigma$ level than the value obtained from 
recent deuterium abundance and microwave background radiation measurements. 
The linear regression slope $dY/d$O = 4.3 $\pm$ 0.7 (corresponding to
$dY/dZ$ = 2.8 $\pm$ 0.5) for the whole sample
of 82 H {\sc ii} regions is similar to that derived for the 
subsample of 7 H {\sc ii} regions, although it has a considerably smaller
uncertainty. 
\end{abstract}


\section{INTRODUCTION}

     It is now well established that four light isotopes, D, $^3$He, $^4$He 
and $^7$Li, were produced by nuclear reactions within the first few minutes 
after the birth of the Universe \citep{R94,WR94,S96,Ty00}. 
In the standard theory of big bang nucleosynthesis (SBBN), given the number 
of light neutrino species $N_\nu$ = 3, 
the abundances of these light elements depend on one cosmological parameter 
only, the baryon-to-photon number ratio $\eta$, which in turn is directly 
related to the density of ordinary baryonic matter $\Omega_bh^2$ \citep{W91},
where $h$ is the Hubble constant in units of 100 km s$^{-1}$ Mpc$^{-1}$. Thus 
precise abundance measurements of the four light elements can provide not 
only a stringest test of the consistency of SBBN, but also information 
about the mean density of ordinary  matter in the Universe.

Deuterium is the best element for deriving the baryonic mass fraction
because its abundance is strongly dependent on $\eta$. Much progress has been
achieved during the last years in the precise measurement of the deuterium
abundance in high-redshift low-metallicity Ly$\alpha$ absorption
systems \citep{BT98a,BT98b,O01,PB01,K03}. These measurements appear to 
converge to the mean value D/H $\sim$ 3 $\times$ 10$^{-5}$
which corresponds to $\Omega_bh^2$ $\sim$ 0.020 $\pm$ 0.002. This value
is in good agreement with the ones of 0.021 -- 0.022 from recent studies of
the fluctuations of the cosmic microwave background (CMB) 
\citep{Pr02,N02,S03}.

     Determining the primordial abundance of $^3$He is
more difficult. Not only it is destroyed in stars, but it
can also be produced by low-mass stars. Thus the derivation of its primordial
value is complicated by our lack of understanding of both the
chemical evolution of the Galaxy and the production of $^3$He in stars.
However, recently \citet{BRB02} determined an upper limit for the primordial 
abundance of $^3$He relative to hydrogen $^3$He/H = (1.1 $\pm$ 0.2) $\times$
10$^{-5}$ by arguing that most solar-mass stars do not produce enough 
$^3$He to enrich 
the interstellar medium significantly. This corresponds to
$\Omega_bh^2$ $\sim$ 0.020$^{+0.007}_{-0.003}$, 
in excellent agreement with the value obtained from 
the deuterium and CMB measurements. 

   As for the primordial abundance of $^7$Li, possible correlations of its 
value with temperature and
metallicity in old hot population II stars may introduce systematic errors.
The value for the $^7$Li primordial abundance derived from the 
$^7$Li abundance plateau of halo stars by \citet{BM97} is $^7$Li/H = 
(1.75$\pm$0.05$_{1\sigma}$$\pm$0.20$_{sys}$)$\times$10$^{-10}$, 
corresponding to two possible values for the baryonic density
$\Omega_bh^2$ = 0.006 and $\Omega_bh^2$ = 0.015, below the value derived
from the D abundance and CMB measurements. Furthermore, \citet{R99} found
a correlation of the $^7$Li abundance with the metallicity of halo stars, and
inferred $^7$Li/H = 1.0 $\times$ 10$^{-10}$ corresponding to a single
value of $\Omega_bh^2$ = 0.009, while \citet{R00} find 
$^7$Li/H = 1.23$^{+0.68}_{-0.32}$ $\times$ 10$^{-10}$, significantly
below the value of 4.5$^{+0.9}_{-0.8}$ $\times$ 10$^{-10}$, predicted 
by SBBN from
the primordial D abundance \citep{K03}.  However, recently \citet{F02} has
shown that the
$^7$Li abundance derived in stars depends on the Li {\sc i} line used.
From the weak Li {\sc i} 
$\lambda$6104 subordinate line instead of the commonly used Li {\sc i} 
$\lambda$6708 resonance line, they derived a much higher $^7$Li abundance
of $\sim$ 3 $\times$ 10$^{-10}$, consistent with the SBBN predicted value.

The primordial mass fraction $Y_p$ of $^4$He can be derived with a much 
better 
precision compared to the primordial abundances of other light elements. $Y_p$
is usually derived by extrapolating the $Y$ -- O/H and
$Y$ -- N/H correlations to O/H = N/H = 0, as proposed originally by 
\citet{PTP74,PTP76} and \citet{P86}. Many attempts at determining $Y_p$ have 
been made, constructing these correlations for
various samples of dwarf irregular and blue compact galaxies (BCGs) (see
references in \citet{IT98b}, hereafter IT98). 
These dwarf systems are the least chemically
evolved galaxies known, so they contain very little helium manufactured by
stars after the big bang, allowing us to bypass the chemical evolution
problems which plague the determination of $^3$He. 
However, because the big-bang production
of $^4$He is relatively insensitive to the density of 
baryonic matter, the primordial
abundance of $^4$He needs to be determined  to a very high precision 
(to better
than 1\%) in order to put useful constraints on $\Omega_b$
and $N_\nu$. Uncertainties in the determination of $Y_p$ can be statistical
or systematic. Statistical uncertainties can be decreased by obtaining
very high signal-to-noise ratio spectra of BCGs. These
BCGs are undergoing intense bursts of star formation, giving birth to high
excitation supergiant H {\sc ii} regions and allowing an accurate determination of the
helium abundance in the ionized gas through the BCG emission-line spectra.
The theory of nebular emission is well-understood enough 
not to introduce additional uncertainty.

Care should also be exercised to consider all possible systematic effects.
We have considered several systematic effects in our previous primordial
helium work. First, we have solved consistently for the electron density
$N_e$(He {\sc ii}) in the He$^+$ zone rather than just setting it to
$N_e$(S {\sc ii}) as was done by previous investigators. Second, we have
corrected the He {\sc i} emission lines for collisional and fluorescent
enhancements. With those systematic effects taken into account,
\citet{ITL97} (hereafter ITL97), by fitting linear regression lines to the
$Y$ -- O/H and $Y$ -- N/H correlations for a sample of 23 BCGs, have 
derived $Y_p$ = 0.243 $\pm$ 0.003.
IT98 by extending that
sample to 45 H {\sc ii} regions in 41 BCGs have obtained 
0.244 $\pm$ 0.002. These values are significantly higher than
those of 0.228 -- 0.234 obtained by \citet{P92} and \citet{O97} with
lower signal-to-noise ratio spectra, by setting $N_e$(He {\sc ii}) =
$N_e$(S {\sc ii}) and not taking into account fluorescent enhancements of
the He {\sc i} lines.

We have also determined $Y$ in individual very metal-deficient BCGs.
\citet{I97}, \citet{I99}, \citet{TIF99}, \citet{G01}, \citet{ICG01} 
have respectively derived the helium abundance in 
the BCGs I Zw 18 ($Z_\odot$/50), SBS 0335--052 
($Z_\odot$/40),
SBS 0940+544 ($Z_\odot$/29), Tol 1214--277 ($Z_\odot$/24), Tol 65 
($Z_\odot$/24) and SBS 1415+437 ($Z_\odot$/21), using high signal-to-noise 
ratio spectra obtained with the MMT and the Keck telescope. Except for
I Zw 18, these determinations have all
resulted in a $^4$He mass fraction in the narrow range $Y$ = 0.245 -- 0.246.
The helium mass fraction for the SE component of I Zw 18, $Y$ = 0.243,
is lower because of the contribution of underlying stellar He {\sc i}
absorption \citep{IT98a,I99}. Using the two
most metal-deficient BCGs known, I Zw 18 and SBS 0335--052, 
\citet{I99} derived a primordial helium mass fraction 
$Y_p$ = 0.245 $\pm$ 0.002.

However, there are other systematic effects which may either increase or
decrease $Y_p$, that were not taken into account in our previous work. 
First, there is the possible
systematic effect of underlying He {\sc i} stellar absorption (ITL97, IT98). 
Correction for this increases $Y$ in the galaxy. 
In particular, He {\sc i} stellar
absorption is important in I Zw 18 \citep{IT98a,I99}. Because it has the
lowest metallicity known, I Zw 18 plays an important role in the determination
of $Y_p$. The lower $Y_p$ derived by other groups \citep{P92,O97} is largely
the result of neglecting underlying He {\sc i} stellar absorption in I Zw 18.
A second possible systematic effect is the collisional excitation of hydrogen
lines \citep{SI01,P02}. Correction for this effect also increases $Y$.
The third systematic effect concerns the temperature structure of the 
H {\sc ii} region.
The electron temperature of the He$^+$ zone $T_e$(He {\sc ii}) is 
usually set equal to the temperature $T_e$(O {\sc iii}) 
derived from the collisionally excited [O {\sc iii}] emission lines.
Recent work has shown that $T_e$(He {\sc ii}) may be smaller than 
$T_e$(O {\sc iii}) \citep{P00,P02,SJ02}. Correction for this effect
decreases $Y$. A last systematic effect concerns the ionization
structure of the H {\sc ii} region. The He$^+$ zone can be larger or smaller
than the H$^+$ zone. Therefore, an ionization correction factor
$ICF$(He) should be applied. In high excitation low-metallicity
H {\sc ii} regions, the ionization correction factor $ICF$(He) may be slightly
higher or lower than unity \citep{V00,ICG01,G02,P02,SJ02}. These different 
systematic effects work in opposite directions
and it is a priori not clear whether combining all of them would increase or
decrease $Y_p$.
Preliminary estimates by \citet{TI02} suggest that, after correction 
for all the systematic effects mentioned above, $Y_p$ may increase by as much
as $\sim$2--4\%.

Besides the $Y_p$ problem, there is also a need to improve the accuracy of 
the determination of the slope $dY/dZ$ from the $Y$ -- O/H linear 
regression.
ITL97 and IT98 have derived $dY/dZ$ =
1.7 $\pm$ 0.9 and 2.4 $\pm$ 1.0 respectively. These values are in agreement, 
within the errors, with the slope $dY/dZ$= 2.1 $\pm$ 0.4 derived by 
\citet{J03} using K dwarf stars with accurate
spectroscopic metallicities in the {\sl Hipparcos} catalog, 
and with the slope $dY/dZ$= 1.5 $\pm$ 0.3 derived by
\citet{P03} from observations of two relatively metal-rich H {\sc ii}
regions, 30 Dor and NGC 346.
They are also in agreement with the values predicted by chemical evolution
models of dwarf galaxies \citep{C95,C99}, although the errors on $dY/dZ$
are still large and do not allow to strongly constrain these models. 
The reason for this state of affair is that, for the $Y_p$ problem,
attention has been paid mainly to very low metallicity BCGs.
For example, the sample
of IT98 contains very few intermediate-metallicity BCGs.

We address the two problems of systematic effects on 
the determination of $Y_p$ and of improving
the accuracy of $dY/dZ$ in this paper. To obtain a more accurate $dY/dZ$, we
have added to our sample 31 new BCGs spanning a considerably wider
range of oxygen abundance, resulting in a total sample of 76 BCGs with 
82 H {\sc ii} regions. To take into account systematic effects on the
determination of $Y_p$, we have studied them in detail in a subsample of 
7 H {\sc ii} regions and then have applied the net correction to the total
sample of 82 H {\sc ii} regions to derive the primordial helium abundance.

     We describe the sample, observations and data reduction in \S2. 
In \S3 we determine the
physical conditions and heavy element abundances for all 82 H {\sc ii}
regions in the total sample. Helium 
abundances are derived in \S4. 
In \S5, we present our new best values for
$Y_p$ and for the linear regression slope $dY/dZ$ which has considerably
reduced errors. In \S3, 4 and 5, to compare the $Y_p$ derived from
our new increased sample of 82 H {\sc ii} regions with our previous
$Y_p$ determinations from smaller samples (ITL97, IT98),
we have proceeded in the same manner as in our previous work concerning
systematic effects. We have determined self-consistently $N_e$(He {\sc ii})
and corrected the He {\sc i} line fluxes for collisional and fluorescent
enhancements. To estimate the net change incurred
by $Y_p$ if all known systematic effects are taken into account, we have
analyzed a restricted subsample of 7 H {\sc ii} regions. We apply the
correction derived from the restricted sample to the whole sample
and present our best value
of $Y_p$ and $dY/dZ$ corrected for systematic effects in \S6.
The implications of our results on cosmology are discussed in \S7. 
We summarize our conclusions in \S8.

\section{OBSERVATIONS AND DATA REDUCTION}

We have obtained new high signal-to-noise ratio spectrophotometric 
observations for 33 H {\sc ii} regions in 31 BCGs with
the Kitt Peak 4-meter telescope on the nights of 2002 December 30 -- 31
and 2003 January 1.
The majority of the galaxies were selected from the First Byurakan 
(Markarian) \citep{M67}, the University of Michigan \citep{SM88}
and Hamburg/SAO \citep{U99} objective prism surveys.
In choosing the candidates, we have used two main selection criteria.
First, they need to have high equivalent widths (EW) 
of the H$\beta$ emission line 
(in general, EW(H$\beta$) $>$ 200\AA) to minimize the influence of underlying 
He absorption on the He abundance determination. Second, they are to span a
wide range of metallicities, from $\sim$ $Z_\odot$/50 to $\sim$ $Z_\odot$/3,
so as to allow an improved determination of $dY/dZ$.
The observed galaxies are listed in Table \ref{tab1} in order of increasing
right ascension, along 
with some of their general properties such as coordinates, apparent
magnitudes, redshifts and absolute magnitudes. Relevant references are
also given.

   All observations were made with the Ritchey-Chr\'etien spectrograph used
in conjunction with the 2048$\times$2048 CCD detector. We use a 
2$''$$\times$300$''$ slit with the grating KPC-10A in first
order, and with a GG 375 order separation filter. This filter cuts off all 
second-order contamination for wavelengths blueward of 7400\AA, which is the 
wavelength region 
of interest here. The above instrumental set-up gave a spatial scale
along the slit of 0\farcs69 pixel$^{-1}$, a scale perpendicular to the slit
of 2.7\AA\ pixel$^{-1}$, a spectral range of 3500--7500\AA\ and a spectral
resolution of $\sim$ 7\AA\ (FWHM). These parameters permitted to cover 
simultaneously the blue and red spectral range with all the lines of interest 
in a single exposure and with enough spectral resolution to separate 
important emission lines as H$\gamma$\ $\lambda$4340 and [O {\sc iii}] 
$\lambda$4363, and H$\alpha$\ $\lambda$6563 and [N {\sc ii}] $\lambda$6584. 
The seeing was in the range 1\arcsec--2\arcsec. Total exposure times varied 
between 20 and 60 minutes. Each exposure was broken up into 2--4 
subexposures, not exceeding 20 minutes, to allow for removal of cosmic rays. 
     Three Kitt Peak IRS spectroscopic standard stars Feige 110, Feige 34 and 
HZ 44 were observed at the beginning, middle and end of each night
for flux calibration. Spectra of He--Ne--Ar comparison arcs were obtained
before or after each observation to calibrate the wavelength scale. The log
of the observations is given in Table \ref{tab2}.

    The two-dimensional spectra were bias subtracted and flat-field corrected
using IRAF\footnote{IRAF is distributed by National Optical Astronomical 
Observatory, which is operated by the Association of Universities for 
Research in Astronomy, Inc., under cooperative agreement with the National 
Science Foundation.}. We then use the IRAF
software routines IDENTIFY, REIDENTIFY, FITCOORD, TRANSFORM to 
perform wavelength
calibration and correct for distortion and tilt for each frame. 
One-dimensional spectra were then extracted from each frame using the APALL 
routine. Before extraction, distinct two-dimensional 
spectra of the same H {\sc ii} region
were carefully aligned using the spatial locations of the brightest part in
each spectrum, so that spectra were extracted at the same positions in all
subexposures. For all objects we extracted the 
brightest part of the BCG, corresponding to a different spatial size
for each object. In two BCGs, CGCG 007--025 and Mrk 450,
spectra of two different H {\sc ii} regions in the same galaxy were extracted.
The extraction apertures of the one-dimensional spectra used for
abundance determinations are given in Table \ref{tab2}. These spectra have
been corrected for the night sky absorption bands in the red part using
spectra of the standard star Feige 34 which contains no absorption lines
in that part of the spectrum. 
All extracted spectra from the same object were then co-added. To perform 
the co-adding and removal of the cosmic-ray hits,
the IMCOMBINE routine is generally used. However,
the use of the IMCOMBINE routine can introduce spurious changes in the line 
intensities of sharp narrow emission lines when small spatial shifts are 
still present. In those cases, we have simply summed the individual
subexposures. Cosmic rays hits have been manually removed in each individual 
subexposure. Then spectra obtained from each individual subexposure
were checked for cosmic rays hits at the location of strong emission lines. 
Fortunately, none of them were found.

Particular attention was paid to the derivation of the sensitivity curve. 
It is crucial that the
latter be obtained with a very high accuracy for a precise
primordial helium abundance determination. To derive the sensitivity curve, we
have fitted with a high-order polynomial the observed spectral energy 
distribution of the bright hot white dwarf standard stars Feige 110, Feige 34 
and HZ 44. Because the spectra of these stars have only a small number of a 
relatively weak absorption features, their spectral energy distributions are 
known with very good accuracy \citep{O90}. 
Moreover, the response function of the CCD detector is smooth, so we could
derive a sensitivity curve with a precision better than 1\% over the
whole optical range, except for the region blueward of [O {\sc ii}] 
$\lambda$3727 where the sensitivity drops precipitously. 
The resulting spectra for all 33 H {\sc ii} regions are shown in 
Figure \ref{fig1}. These spectra have been reduced to zero redshift and 
corrected for interstellar extinction.

     As a whole, the H {\sc ii} regions in our new 
BCG sample show high excitation
spectra, except for Mrk 1063. Nineteen H {\sc ii} regions have a H$\beta$ 
equivalent width greater than 200\AA.
   The observed line fluxes $F$($\lambda$) normalized to $F$(H$\beta$) and
their errors for the 33 H {\sc ii} regions shown in Figure \ref{fig1} are 
given in Table \ref{tab3}. They were measured using the IRAF SPLOT routine.
The line flux errors listed include statistical errors derived with
SPLOT from non-flux calibrated spectra, in addition to errors introduced
in the standard star absolute flux calibration, which we set to 1\% of the
line fluxes. These errors will be later propagated into the calculation
of abundance errors.
The line fluxes were corrected for both reddening \citep{W58} 
and underlying hydrogen stellar absorption derived simultaneously by an 
iterative procedure as described in \citet{ITL94} (hereafter ITL94). The 
corrected line 
fluxes $I$($\lambda$)/$I$(H$\beta$), equivalent widths EW($\lambda$), 
extinction
coefficients $C$(H$\beta$), and equivalent widths EW(abs) of the hydrogen
absorption stellar lines are also given in Table \ref{tab3}, along with the 
uncorrected H$\beta$ fluxes. 

\section{PHYSICAL CONDITIONS AND DETERMINATION OF HEAVY ELEMENT ABUNDANCES}

   To determine element abundances, we follow 
the procedures of ITL94, ITL97 and
\citet{TIL95}. We adopt a two-zone photoionized H {\sc ii}
region model: a high-ionization zone with temperature $T_e$(O {\sc iii}), 
where O {\sc iii}, Ne {\sc iii} and Ar {\sc iv} lines originate, and a 
low-ionization zone with temperature $T_e$(O {\sc ii}), where O {\sc ii}, 
N {\sc ii}, S {\sc ii} and Fe {\sc iii} lines originate. 
As for the Ar {\sc iii} and S {\sc iii} lines they originate in the 
intermediate zone between the high and low-ionization regions \citep{G92}. 
We have derived the chlorine abundance from Cl {\sc iii} emission lines in 22 
H {\sc ii} regions. We assume that these lines also originate in the 
intermediate zone as 
the ionization potentials of the S$^{++}$, Ar$^{++}$ and Cl$^{++}$ ions are 
similar. The temperature $T_e$(O {\sc iii}) is calculated using the 
[O {\sc iii}] $\lambda$4363/($\lambda$4959+$\lambda$5007) ratio. To take into 
account the electron temperatures for different ions, we have used 
the H {\sc ii} photoionization models of \citet{S90}, as fitted by 
the expressions in ITL94 and ITL97. The [S {\sc ii}] 
$\lambda$6717/$\lambda$6731 ratio is used to determine the electron density 
$N_e$(S {\sc ii}), the minimum value of which was set to be 10 cm$^{-3}$.
Ionic and total heavy element abundances are derived for the 33 H {\sc ii} 
regions in the manner described in ITL94 and are given in 
Table \ref{tab4} along with the adopted ionization correction factors ($ICF$).

As a result of our sample selection criteria,
the oxygen abundances in the newly observed H {\sc ii} regions span a wide
range of metallicities, going from 12 + log (O/H) = 7.43 
($Z_\odot$/30, J 0519+0007) to 8.30 ($Z_\odot$/4, Mrk 35) with a large
fraction of high-metallicity galaxies. Five galaxies among those observed have
12 + log (O/H) $>$ 8.2 ($Z_\odot$/5). This large metallicity range is crucial
for determining an accurate $dY/dZ$ slope.

\section{HELIUM ABUNDANCE}

     Because the $Y_p$ -- $\eta$ relation has a very shallow slope, 
$Y_p$ has to
be determined to better than 1\% in order to put interesting 
constraints on the baryonic mass fraction of the universe.
In principle, this high precision can be achieved if care is taken to: 
1) obtain spectra of low-metallicity H {\sc ii} regions with the very
highest signal-to-noise ratio; 2) take into account all physical processes
which may make He {\sc i} line intensities deviate from their recombination 
values such as collisional and fluorescence enhancements; 3) use the best 
possible atomic data; and 4) take into account systematic effects in the 
determination of the
helium mass fraction, such as underlying stellar absorption, the 
ionization and temperature structures of the H {\sc ii} region.

In this section, to compare the results from the present enlarged sample
with our previous work, we derive $Y_p$ in the same way as
ITL94, ITL97 and IT98. We solve self-consistently for $N_e$(He {\sc ii})
and the optical depth $\tau$($\lambda$3889) in the He {\sc i} 
$\lambda$3889 emission line, but do not consider possible temperature and
ionization structures in the H {\sc ii} regions. Thus, we set
the temperature in the He$^+$ zone to be equal to that derived from the
[O {\sc iii}] $\lambda$4363/($\lambda$4959 + $\lambda$5007) flux ratio, i.e.
$T_e$(He {\sc ii}) = $T_e$(O {\sc iii}). The case 
where $T_e$(He {\sc ii}) $\leq$ 
$T_e$(O {\sc iii}) will be discussed in \S6 for a subsample of 7 H {\sc ii}
regions. The He ionization correction factors are set to 1.
We consider collisional and fluorescence enhancement effects but
do not take into account other systematic effects such as 
the collisional excitation of hydrogen lines and underlying
He {\sc i} stellar absorption. These effects will be discussed in \S6 as well.

Because observed He {\sc i} fluxes
deviate from their recombination values, corrections are necessary to
take into account
the physical mechanisms which modify line fluxes. 
In the range of high temperatures found in BCGs
(10$^4$K $\la$ $T_e$ $\la$ 2$\times$10$^4$K), the main such physical 
mechanism is collisional excitation from the metastable 2$^3$S level 
of He {\sc i}. The second physical mechanism to consider is fluorescence 
enhancement where self-absorption in some optically thick emission lines 
populates the upper levels of He {\sc i}. 
To correct line intensities for these effects we have evaluated the
electron number density $N_e$(He {\sc ii}) and the optical depth 
$\tau$($\lambda$3889) in the He {\sc i} $\lambda$3889 line in a 
self-consistent way by minimization of the expression
\begin{equation}
\chi^2=\sum_i^n\frac{(y^+_i-y^+_{\rm mean})^2}{\sigma^2(y^+_i)}\label{eq1},
\end{equation}
where $y^+_i$ is the He$^+$ abundance derived from the flux of the He {\sc i}
emission line with label $i$. The quantity $y^+_{\rm mean}$ is the weighted 
mean of the He$^+$
abundance as derived from the equation
\begin{equation}
y^+_{\rm mean}=\frac{\sum_i^k{y^+_i/\sigma^2(y^+_i)}}
{\sum_i^k{1/\sigma^2(y^+_i)}}\label{eq2}.
\end{equation}
The numbers of lines $n$ and $k$ used for the minimization and determination
of the weighted mean can be different, and in general $n \geq k$.

In this paper, following ITL94, ITL97 and IT98 we use the five 
strongest He {\sc i} $\lambda$3889, $\lambda$4471, 
$\lambda$5876, $\lambda$6678 and $\lambda$7065 emission lines to derive 
$N_e$(He {\sc ii}) and $\tau$($\lambda$3889), i.e. $n = 5$.
The He {\sc i} $\lambda$3889 and 
$\lambda$7065 lines play an important role because they are particularly 
sensitive to both quantities. Since the
He {\sc i} $\lambda$3889 line is blended with the H8 $\lambda$3889 line, 
we have subtracted the latter, assuming its intensity to be equal to 0.107 
$I$(H$\beta$) \citep{A84}. However, the He$^+$ abundances derived from the
He {\sc i} $\lambda$3889 and $\lambda$7065 emission lines 
are more uncertain as compared to
those derived from other He {\sc i} emission lines because of their higher 
sensitivity to collisional and fluorescent enhancements and, in the case
of $\lambda$3889 line, 
also because of the uncertainties due to its blending with 
the H8 line. Therefore, we will consider two cases, one with $k$ = 5 when all 
lines are used in the calculation of $y^+_{\rm mean}$, and the other with
$k$ = 3 when He {\sc i} $\lambda$3889 and $\lambda$7065 are excluded.

In our spectra, other He {\sc i} emission lines are seen, most often
He {\sc i} $\lambda$3820, $\lambda$4387, $\lambda$4026, $\lambda$4921,
$\lambda$7281. However, we do not attempt to use these lines for He 
abundance determination because they are much weaker as compared to the five 
brightest lines, and hence have larger uncertainties 
(Table \ref{tab3}).

Concerning the atomic data, we have used in our previous work
(ITL97, IT98, \citet{I99}, \citet{G01}, \citet{ICG01}) 
the He {\sc i} emissivities of 
\citet{Sm96} along with He {\sc i} collisional enhancement
coefficients of \citet{KF95a}, and fits of the
fluorescent enhancement factors of \citet{R68} by ITL97. Recently, 
\citet{B02} have computed new He {\sc i} emission line 
strengths taking into account both collisional and fluorescent enhancements.
Their calculations are also based on  \citet{Sm96}' atomic data. 
They give simple and convenient fits for calculations of abundances for
different He {\sc i} emission lines.
These authors also show 
that some of the correction factors used by ITL97 may be in error
by a factor of $\sim$2, especially for 
He {\sc i} $\lambda$7065 emission line, which may lead to an overestimation
of $Y_p$. Thus, in this paper we use the new
\citet{B02} fits to convert He {\sc i} emission-line strengths to singly 
ionized helium $y^+$ $\equiv$ He$^+$/H$^+$ abundances.

   Additionally, in those cases when the nebular He {\sc ii} $\lambda$4686 
emission line was detected, we have added the abundance of doubly ionized 
helium $y^{++}$ $\equiv$ He$^{++}$/H$^+$ to $y^+$. The value of $y$$^{++}$ is 
small ($\leq$ 3\% of $y^+$) in all cases.

   Finally the helium mass fraction is calculated as
\begin{equation}
Y=\frac{4y[1-20({\rm O/H})]}{1+4y},                     \label{eq:Y}
\end{equation}
where $y$ = $y^+$ + $y^{++}$ \citep{P92}.

In Table \ref{tab4}, we give for each galaxy
$y^+$($\lambda$4471), $y^+$($\lambda$5876), $y^+$($\lambda$6678), 
$y^{++}$($\lambda$4686), the weighted mean helium abundance $y$ and the
weighted mean helium mass
fraction $Y$. The weighted mean values are derived using only the three
He {\sc i} $\lambda$4471, $\lambda$5876 and $\lambda$6678 emission lines.

\section{THE $dY/dZ$ SLOPE}

   To determine $Y_p$ and $dY/dZ$, we fit the data points in the 
$Y$ -- O/H and $Y$ -- N/H planes by linear regression lines of the
form \citep{PTP74,PTP76,P92}
\begin{equation}
Y = Y_p + \frac{dY}{d({\rm O/H})} ({\rm O/H}),               \label{eq:YvsO}
\end{equation}
\begin{equation}
Y = Y_p + \frac{dY}{d({\rm N/H})} ({\rm N/H}).               \label{eq:YvsN}
\end{equation}

     The sample used to determine $Y_p$ and $dY/dZ$ is composed of 89 
different observations
of 82 H {\sc ii} regions in 76 BCGs. They are listed in Table \ref{tab5}
in order of increasing ionized gas metallicity. 
In addition to the new data for the
33 H {\sc ii} regions presented here (Table \ref{tab1}) we have also
included the 45 H {\sc ii} 
regions from IT98, the 2 H {\sc ii} regions in Tol 1214--277 and 
Tol 65 from \citet{ICG01}, the H {\sc ii} region in HS 1442+4250 from 
\citet{G03a} and H {\sc ii} region No. 2 in SBS 1415+437 from \citet{G03b}.
The number of data points (89) is larger than the number of H {\sc ii} 
regions (82) because several H {\sc ii} regions have independent observations
from different telescopes. We treat these independent observations as separate
data points in our least-square fitting.
Thus, I Zw 18 was observed with the 4m KPNO telescope and the MMT, 
SBS 0335--052 with the 2.1m KPNO telescope and with 
Keck (twice), SBS 0940+544 with the 4m KPNO telescope, the MMT and Keck,
SBS 1415+437(\#1) with the MMT and the 4m KPNO telescope (twice).

In order to have a homogeneous sample with all spectra reduced and analyzed 
in exactly the same way, we have recalculated the He abundances of all our
previously published objects with the new \citet{B02} 
corrections for collisional and fluorescent enhancements.

In Figure \ref{fig2} we show by solid lines the $Y$ -- O/H and $Y$ -- N/H 
regression lines for 
the whole sample. The dashed lines are 1$\sigma$ alternatives. 
The IT98 data are represented by open circles, other
 data from \citet{I99} (I Zw 18 and SBS 0335--052), \citet{TIF99} 
(SBS 1415+437\#1), \citet{ICG01} (Tol 1214--277 and Tol 65), \citet{G01}
(SBS 0940+544), \citet{G03a} (HS 1442+4250), \citet{G03b} (SBS 1415+437\#1 and
SBS 1415+437\#2) by stars, and the new data presented 
in this paper by filled circles. Note that the sample includes now,
in contrast to our previous work (ITL97, IT98),
a significant number of the galaxies with relatively high oxygen abundances.

In Figs. \ref{fig2}a -- \ref{fig2}b the He abundances have been obtained using
all 5 lines for $\chi^2$ minimization, but only 3 lines for abundance 
calculations
($n$ = 5, $k$ = 3 in Eqs. \ref{eq1} and \ref{eq2}). In Figs. \ref{fig2}c -- 
\ref{fig2}d all 5 lines are used for both $\chi^2$ minimization and abundance
calculations ($n$ = $k$ = 5 in Eqs. \ref{eq1} and \ref{eq2}). It can be seen
that there is no systematic difference between the old and new data in 
Figs. \ref{fig2}a -- \ref{fig2}b, while the old data have systematically 
lower He abundances as compared to the new data in 
Figs. \ref{fig2}c -- \ref{fig2}d. This is because the He abundances
derived from the He {\sc i} $\lambda$3889 emission line, included in the
second case but not in the first, are systematically lower.
The He {\sc i} $\lambda$3889 line
is subject to large uncertainties introduced by the corrections for hydrogen 
nebular emission and stellar absorption,
as discussed by \citet{OS01} using the data of
IT98. It is likely that the contribution of hydrogen absorption
has been underestimated in our analysis.
The effect is higher in objects with lower equivalent
widths of emission lines since the relative contribution of underlying 
absorption is 
larger. This is the case for the IT98 galaxies. They generally have
lower equivalent widths of the H$\beta$ emission line as compared 
to the galaxies in the new sample, chosen to have large 
EW(H$\beta$). Hence the He {\sc i} $\lambda$3889 line in the new galaxies
is less affected by stellar absorption and thus their derived $Y$s are 
systematically higher than those of the IT98 galaxies. 
The He abundance derived from the He {\sc i} $\lambda$7065 emission line
is also uncertain because of its high sensitivity to collisional and 
fluorescent enhancements. Therefore, because they are less influenced
by the aforementioned effects, the He abundances derived by using only 
the three He {\sc i} $\lambda$4471, $\lambda$5876 and $\lambda$6678
emission lines for the IT98 galaxies are much more consistent with those
for the new galaxies.

The parameters of the linear regression
fits for the old and (old+new) samples are given in Table \ref{tab6}. 
We also show the dispersions $\sigma$ of $Y$ about the regression lines.
The first line of Table \ref{tab6} gives the parameters for the IT98 
sample of 45 H {\sc ii} regions, for which we have recalculated He 
abundances with the \citet{B02} equations. The primordial He 
abundance derived from the $Y$ -- O/H relation is 
$Y_p$ = 0.245 $\pm$ 0.002, slightly larger than the value of 0.244 $\pm$ 0.002
obtained by IT98 for the same sample with
the ITL97 and IT98 expressions for collisional and 
fluorescent enhancements of He {\sc i} emission lines. 
The higher $Y_p$ value is mainly the consequence of a slope which is $\sim$2
times shallower ($dY$/$d$(O/H) = 21 $\pm$ 21 in Table \ref{tab6} as compared
to 45 $\pm$ 19 in IT98).
In the second line are given the regression parameters for the total sample. 
From the $Y$ -- O/H relation, we obtain $Y_p$ = 0.243 $\pm$ 0.001, 
slightly lower
than the value derived from the IT98 sample, because of the steeper slope
($dY$/$d$(O/H) = 51 $\pm$ 9 as compared to 21 $\pm$ 21).
The value derived from the $Y$ -- N/H linear regression, is slightly higher,
$Y_p$ = 0.244 $\pm$ 0.001. This is because the very
highest metallicity BCGs in our sample have a slightly higher N/O ratio than
the lower metallicity galaxies, thus making the $Y$ -- N/H relation slightly
non-linear. This higher N/O ratio is caused by the presence of secondary
nitrogen in the higher-metallicity BCGs. The standard deviation of the data
points is 0.004 for both $Y$ -- O/H and $Y$ -- N/H regressions, lower than
for the IT98 sample. Lines 3 and 4 of Table \ref{tab6} will be discussed in 
\S6.

The slopes of the $Y$ -- O/H and $Y$ -- N/H linear regressions can be 
written as:

\begin{equation}
\frac{dY}{d({\rm O/H})} = 12\frac{dY}{d{\rm O}} = 
18.2\frac{dY}{dZ}, \label{eq:dO}
\end{equation}
\begin{equation}
\frac{dY}{d({\rm N/H})} = 10.5\frac{dY}{d{\rm N}} = 
564\frac{dY}{dZ}, \label{eq:dN}
\end{equation}
where O, N and $Z$ are respectively the mass fractions of oxygen, nitrogen 
and heavy elements. We have assumed that O = 0.66$Z$ \citep{M92} for a 
metallicity $Z$=0.001, an IMF slope $x$=1.35 (where $x$ is defined by 
$dN/d({\rm log}M)$ $\propto$ $M^x$) and log(N/H) -- log(O/H) = --1.55 
\citep{TIL95}. 

   From Table \ref{tab6} we derive $dY/d{\rm O}$ = 4.3 $\pm$ 0.7
and $dY/dZ$ = 2.8 $\pm$ 0.5. Although the slope $dY/dZ$ derived here is 
consistent within the errors with $dY/dZ$ = 2.6 $\pm$ 1.4 and 3.7 $\pm$ 1.5
derived respectively by ITL97 and IT98, we have decreased the errors on the
slope by a factor of $\sim$ 3 thanks to the larger number of high metallicity
objects included in the sample. 
Our new value of $dY/dZ$ is also consistent with that
of 2.1 $\pm$ 0.4 derived by \citet{J03} from nearby K dwarf stars.
The extrapolation of the oxygen linear regression with 
$dY/d{\rm O}$ = 4.3 to solar metallicity gives $Y$ = 0.274, in good 
agreement with $Y$ = 0.271 derived for the Sun \citep{C98}.

The increased accuracy of our $dY/d$O and $dY/dZ$ values 
allows us to constrain more
stringently chemical evolution models. Our slope $dY/dZ$ = 2.8 $\pm$ 0.5
appears to be too steep as compared to the values 1.0 -- 1.2 obtained by
\citet{L01} with their closed-box chemical evolution models for gas-rich
dwarf galaxies and their adopted O=0.6$Z$. Open models with ordinary 
non-enriched winds, with or without
gas inflow, predict also the same range of values 1.0 -- 1.2. 
The models of \citet{L01} with
oxygen-enriched winds predict steeper slopes, more in agreement with our
value, but would not fit gas mass fractions and N/O ratios simultaneously.
On the other hand, analytical chemical evolution models with 
selective heavy element loss \citep{P94} predict $dY/d$O $>$ 5 
with $dY/d$O increasing with increasing efficiency of the oxygen-enriched
galactic wind. Hence, our derived $dY/d$O = 4.3 $\pm$ 0.7 excludes
chemical evolution models with a high mass loss of the enriched gas.
Our $dY/d$O is in agreement with the values predicted by
chemical evolution closed-box models and models with
non-selective galactic winds of \citet{C99} for dwarf irregular galaxies. 
However, those multi-parameter models do not consider the evolution of 
some elements, such as nitrogen, and it is not clear therefore, whether they 
can reproduce all the observational constraints for dwarf galaxies.

The above values were derived by neglecting other possible systematic
effects, such as the difference between the electron temperatures in the
He {\sc ii} and O {\sc iii} zones, underlying He {\sc i} stellar absorption, 
the collisional excitation of hydrogen lines, and the ionization structure
of the H {\sc ii} region. In the next section, we investigate, how these
systematic effects may change our derived values of $Y_p$ and $dY/dZ$,
using a subsample of 7 H {\sc ii} regions.

\section{SYSTEMATIC EFFECTS}

We will consider successively the following four systematic effects: 
1) the underlying stellar He {\sc i} absorption; 
2) the collisional excitation of
hydrogen lines; 3) the temperature structure of the H {\sc ii} region,
i.e. the temperature difference between $T_e$(He {\sc ii})
and $T_e$(O {\sc iii}); and 4) the ionization structure of the H {\sc ii}
region.

\subsection{Underlying stellar He {\sc i} absorption} 

It has long been recognized 
\citep{R82,KS83,DS86,P92,O95} that absorption caused by hot stars in 
the He {\sc i} lines can decrease
the intensities of the nebular He {\sc i} lines. In particular, \citet{IT98a}
have shown that the neglect of He {\sc i} underlying stellar absorption has 
led to the derivation of a very low helium mass fraction in I Zw 18, the 
most metal-deficient BCG known \citep{P92,O97}.
This effect is most pronounced 
for the He {\sc i} $\lambda$4471 emission line. Recently, 
\citet{G99} have produced synthetic spectra of H Balmer and He {\sc i}
absorption lines in starburst and poststarburst galaxies. They predict an
equivalent width of the He {\sc i} $\lambda$4471 absorption line 
in the range $\sim$ 0.4 -- 0.8 \AA, or
$\sim$ 10 -- 20\% of the He {\sc i} $\lambda$4471 emission line equivalent
width for 
young starbursts with an age $\la$ 7 Myr, which is the case for 
our H {\sc ii} regions. 
However, the equivalent width of the He {\sc i} $\lambda$3889 absorption 
line is not known. \citet{G99} predict the total 
equivalent width of the H8 and He {\sc i} $\lambda$3889 absorption lines
to be in the range of 0.85 -- 0.95 that of the H$\beta$ absorption line, but 
do not give the absorption equivalent width for the He {\sc i} $\lambda$3889 
line alone.
Those authors did not calculate either absorption line equivalent 
widths for the other prominent He {\sc i} lines, although 
the effect of underlying
absorption appears to be smaller for the He {\sc i} $\lambda$5876,
$\lambda$6678 and $\lambda$7065 emission lines. 
In any case, underlying stellar absorption must be taken into account for
all He {\sc i} lines if we are to achieve the desired high precision of
$\la$1\% in the primordial He abundance determination.

\subsection{Collisional excitation of H lines}

It was generally assumed in abundance studies that deviations of the
observed H$\alpha$/H$\beta$ flux ratio from the theoretical recombination
value are entirely due to interstellar extinction. \citet{DK85} first
noted that in the hot and dense H {\sc ii} regions of BCGs, collisional 
excitation of hydrogen emission lines can be important.
This can affect the derived H abundances and hence the He abundances since the
latter are always derived relative to H abundances. \citet{SI01} estimated
that this effect can result in an upward correction in the He abundances
of up to 5\%, assuming
that the excess of the H$\alpha$/H$\beta$ flux ratio above the
theoretical recombination value is due only to collisional excitation.
Realistically however, the H$\alpha$/H$\beta$ flux ratio excess is 
also partly caused by interstellar extinction, so that 
the effect of collisional excitation of the hydrogen lines is
likely smaller, $\sim$ 2 -- 3 percent, similar to the estimate of \citet{P02}.

\subsection{Temperature structure}

To derive the He abundances, ITL94, ITL97 and IT98 have assumed 
that the temperatures $T_e$(He {\sc ii}) and $T_e$(O {\sc iii}) averaged over
the whole H {\sc ii} region are equal. $T_e$(O {\sc iii}) is 
determined from the observed 
[O {\sc iii}]$\lambda$4363/($\lambda$4959 + $\lambda$5007) emission line flux 
ratio. However, because of the high sensitivity of the flux of the auroral 
[O {\sc iii}]$\lambda$4363 emission line to temperature, 
$T_e$(O {\sc iii}) tends to be characteristic of the hotter zones in the 
H {\sc ii} region, and it is generally higher than $T_e$(He {\sc ii}). To 
take into account the difference of the calculated electron temperature 
$T_e$(O {\sc iii}) 
as compared to the one in the H$^{+}$ and He$^{+}$ zones, \citet{P67} 
developed a formalism based on the average temperature $T_0$ and 
the mean square temperature
variation $t^2$ in an H {\sc ii} region. Then the temperatures 
$T_e$(H {\sc ii}), $T_e$(He {\sc ii}) and $T_e$(O {\sc iii})
are expressed as different functions
of $T_0$ and $t^2$, and $T_e$(O {\sc iii}) $\geq$ 
$T_e$(H {\sc ii}), $T_e$(He {\sc ii}). This approach has been applied 
by \citet{P02} for the determination of the He abundance in some 
low-metallicity dwarf galaxies, including the two most-metal 
deficient BCGs known, I Zw 18 and SBS 0335--052. Using the observations of
\citet{I99} they find that the difference between
$T_e$(O {\sc iii}) and $T_e$(He {\sc ii}) results in a reduction of the
He mass fraction by 2 -- 3 percent as compared to the case when 
$T_e$(O {\sc iii}) = $T_e$(He {\sc ii}).

\subsection{Ionization structure} 

    Another source of systematic uncertainties comes from the assumption
that the H$^+$ and He$^+$ zones in the H {\sc ii} region are coincident. 
However, depending on the hardness of the ionizing radiation, the radius of 
the He$^+$ zone can be smaller than the radius of the H$^+$ zone in
the case of soft ionizing radiation, or larger in the case of hard
radiation. In the former case, a correction for unseen neutral helium
should be made, resulting in an ionization correction factor $ICF$(He) $>$ 1
and hence a higher helium abundance. In the latter case, the situation is 
opposite and $ICF$(He) $<$ 1. The ionization correction factor problem  has 
been discussed in several studies \citep{P92,S97,O97,V00,P00,B00,SJ02}. 
It was shown that the correction of the helium 
abundance can be as high as several percent in either downward or upward
directions depending on the hardness of the radiation and the ionization
parameter $U$. \citet{SJ02} have calculated an
extensive grid of photoionized H {\sc ii} region models aiming to derive the
correction factors as functions of hardness and $U$. Their conclusion was 
that a downward correction of $Y$ by as much as 6\% and 2\% is required 
respectively for ionization parameters log $U$ = --3.0 and --2.5.
However, the downward correction is $\la$ 1\% if log $U$ $\ga$ --2.0, which
is the case for the majority of our H {\sc ii} regions. Therefore, we will
not consider further this effect in the present paper and set $ICF$(He) = 1.

\subsection{Analysis of a subsample of 7 H {\sc ii} regions}

Although these systematic effects are still poorly studied and difficult
to determine precisely, \citet{ICG01} have concluded that when taken 
into account together, they largely offset each other. 
\citet{TI02} have suggested that combining all systematic effects may push
$Y_p$ upward by at most $\sim$ 2 -- 4\%. 
\citet{P02} have considered collisional and 
fluorescent enhancements, along with all systematic effects
discussed in this section, in their analysis of the He abundance
in five H {\sc ii} regions. They attack
the problem in three steps. First, using the ionization code CLOUDY 
\citep{F96,F98} these authors 
estimate 
the parameter $t^2$,
the optical depth 
$\tau$($\lambda$3889), the ionization correction factor $ICF$(He) and the 
contribution of collisional enhancement to the
hydrogen lines. Second, they adopt the equivalent 
widths of the H {\sc i} and He {\sc i} absorption lines calculated by 
\citet{G99} for a particular age of the star formation burst in each
H {\sc ii} region, except for NGC 346 where the equivalent widths of 
absorption lines were set to zero. 
No correction for underlying stellar absorption was applied to the strong 
He {\sc i} $\lambda$5876, $\lambda$6678 and $\lambda$7065 emission lines.
These lines play an important role in the helium abundance determination,
but their absorption equivalent widths were not calculated by \citet{G99}.
\citet{P02} then use a $\chi^2$ minimization 
procedure with input parameters $t^2$, 
$\tau$($\lambda$3889), $T_e$(O {\sc iii}) and the equivalent widths of the
He {\sc i} absorption lines to find new values of $t^2$, 
$\tau$($\lambda$3889) and derive the electron temperature $T_e$(He {\sc ii}),
the electron number density $N_e$(H {\sc ii}) and the He abundance.
In the case of the two most metal-deficient BCGs known, I Zw 18 and 
SBS 0335--052, \citet{P02} derive, after correction for all systematic 
effects, helium abundances very close to the values obtained 
when systematic effects other than collisional and fluorescent 
enhancements are not taken into account. However, for three other 
higher-metallicity H {\sc ii} regions in their sample,
they derive also relatively low He mass 
fractions, similar to those of I Zw 18 and SBS 0335--052. 
This resulted in a very flat $Y$ -- O/H linear regression, leading these
authors to conclude that they have probably overestimated the collisional
excitation of hydrogen lines in I Zw 18 and SBS 0335--052, resulting in
artificially high He mass fractions.

To estimate the systematic effects in our derived $Y_p$ value in \S5,
and to compare our results with those of \citet{P02}, we have analyzed 
the same objects as these authors: I Zw 18 and SBS 0335--052
\citep{I99}, Mrk 209 $\equiv$ Haro 29 and Mrk 71 (ITL97) and 
NGC 346 \citep{P00}. Because the range of
oxygen abundance spanned by the five galaxies is relatively small, we 
have added
two more metal-rich galaxies Mrk 450 (this paper) and UM 311 (IT98)
to improve the metallicity range and obtain a more 
accurate determination of $dY/dZ$. However, our approach is somewhat 
different from that of \citet{P02}.

Instead of using CLOUDY to derive model parameters,
we adopt the self-consistent approach which 
we have used in all of our previous studies and in \S4 of this paper.
Our method consists of using the five strongest He {\sc i} lines 
$\lambda$3889, $\lambda$4471, $\lambda$5876, $\lambda$6678 and 
$\lambda$7065 and correct for systematic effects
in such a way so as to obtain after correction the best agreement between 
the He abundances derived from each He {\sc i} line separately, i.e. we 
search for the minimum $\chi^2$ as defined by
Eq. \ref{eq1}. In our previous work, we have considered as free parameters 
the electron
number density $N_e$(He {\sc ii}) and the optical depth $\tau$($\lambda$3889).
Here we add the electron temperature $T_e$(He {\sc ii}) and 
$\Delta$$I$(H$\alpha$)/$I$(H$\alpha$), the fraction of
the H$\alpha$ flux due to collisional excitation. These values 
should be well determined as we have only 
four unknowns for five constraints, i.e. the problem is overdetermined. 

We proceed in the following manner. We first correct for collisional
excitation of the hydrogen lines by varying 
$\Delta$$I$(H$\alpha$)/$I$(H$\alpha$) in the range between
0 and 5\%, 
and subtracting $\Delta$$I$(H$\alpha$) and $\Delta$$I$(H$\beta$) 
respectively from the observed fluxes of H$\alpha$ and H$\beta$.  We adopt 
$\Delta$$I$(H$\beta$)/$I$(H$\beta$) = 0.33 $\times$ 
$\Delta$$I$(H$\alpha$)/$I$(H$\alpha$) \citep{SI01}.
Then the whole spectrum is corrected for extinction and underlying hydrogen
absorption derived from the hydrogen emission line fluxes, 
including the new reduced H$\alpha$ and H$\beta$ fluxes.

Underlying He {\sc i} absorption constitutes a major 
problem as each of the He {\sc i} 
$\lambda$3889, $\lambda$4471, $\lambda$5876, $\lambda$6678 and 
$\lambda$7065 lines has its own absorption equivalent width EW$_a$. 
Additionally, He {\sc i} $\lambda$3889 is blended with H8 and deblending it
is subject to large uncertainties because of the imperfect correction for 
hydrogen emission and absorption.

Because the equivalent
widths of He {\sc i} absorption lines are dependent on the age of the
star formation burst, we first need to determine the age of the burst using 
the equivalent width of the H$\beta$ emission line. The values
of EW(H$\beta$) for the H {\sc ii} regions considered in this section are 
in the range 100 -- 250\AA, which corresponds to a burst age of
$\sim$4 Myr for the metallicity $Z$ $\la$ $Z_\odot$/5 \citep{SV98}. The
equivalent widths of the He {\sc i} absorption lines have large variations
in the age range 3 -- 5 Myr \citep{G99} and are not well constrained. 
In particular, EW$_a$(He {\sc i} $\lambda$4471) at $Z$ = $Z_\odot$/20 varies 
in the range 0.4 -- 0.6\AA, or by factor of $\sim$ 1.5. As for the equivalent 
width of the blend H8 + He {\sc i} $\lambda$3889, it varies in the range 
2.0 -- 3.9\AA, or by factor of $\sim$ 2. 
The mean equivalent widths in the 3 -- 5 Myr age range 
are EW$_a$(He {\sc i} $\lambda$4471) $\approx$ 0.5\AA\ and 
EW$_a$(H8 + He {\sc i} $\lambda$3889) $\approx$ 3.0\AA.
These values can be smaller 
if the gaseous continuum emission is significant. 
In the H {\sc ii} regions considered in this section, 
the gaseous continuum emission contributes $\sim$ 10\% -- 25\% 
of the total continuum flux at the wavelength of H$\beta$,
reducing by the same 
amount the equivalent width of the H$\beta$ absorption line. However, at 
shorter wavelengths the effect of gaseous continuum is smaller. We thus adopt
two sets of equivalent widths: 
1) EW$_a$(H8 + He {\sc i} $\lambda$3889) = 3.0\AA\ and 
EW$_a$(He {\sc i} $\lambda$4471) = 0.4\AA, which takes into account the 
dilution by the gaseous continuum, and 2) EW$_a$(H8 + He {\sc i} 
$\lambda$3889) = 
3.0\AA\ and EW$_a$(He {\sc i} $\lambda$4471) = 0.5\AA\, where gaseous 
continuum emission is negligible. As for the other He {\sc i} lines,
their EW$_a$ decrease progressively from the blue to the red
(see for example the spectrum of I Zw 18 NW in Fig. 2 of \citet{I99}
which is strongly affected by He {\sc i} underlying absorption and where
He {\sc i} $\lambda$4471 is barely seen while both He {\sc i} $\lambda$6678
and $\lambda$7065 can be seen in emission).
We adopt for them EW(He {\sc i} $\lambda$5876) = 0.3
EW(He {\sc i} $\lambda$4471), EW(He {\sc i} $\lambda$6678) =
EW(He {\sc i} $\lambda$7065) = 0.1 EW(He {\sc i} $\lambda$4471) in both cases.

For each of these two sets of EW$_a$, we
solve the problem self-consistently to find the four parameters 
$\Delta$$I$(H$\alpha$)/$I$(H$\alpha$), $T_e$(He {\sc ii}), 
$N_e$(He {\sc ii}) and $\tau$($\lambda$3889) using the five He {\sc i} 
emission line fluxes. We 
calculate $\chi^2$, 
$T_e$(He {\sc ii}), $N_e$(He {\sc ii}), and $\tau$($\lambda$3889)
as a function of $\Delta$$I$(H$\alpha$)/$I$(H$\alpha$) in the range 0 -- 5\%, 
 in steps of 0.0005, for
a total of 100 models. The best solution corresponds to
the minimum $\chi^2$. It is the one 
in which the separate helium abundances $y$ derived from each individual 
He {\sc i} line are in best agreement with
the weighted mean abundance. 
To exclude unphysical solutions we consider only models with
0.9 $\leq$ $T_e$(He {\sc ii})/$T_e$(O {\sc iii}) $\leq$ 1.0,
10 cm$^{-3}$ $\leq$ $N_e$(He {\sc ii}) $\leq$ 450 cm$^{-3}$ and 
0 $\leq$ $\tau$($\lambda$3889) $\leq$ 5. The values found by
\citet{P02} in the analysis of their H {\sc ii} region sample fall within
those ranges.

The parameters of the best models for each set of EW$_a$s and for each 
of the 7 H {\sc ii} regions are shown in Table 7. The best solutions 
are also shown 
in the $Y$ -- O/H and $Y$ -- N/H planes   
in Fig. \ref{fig3}a -- \ref{fig3}b (models with 
EW$_a$(H8 + He {\sc i} $\lambda$3889) = 3.0\AA\ and 
EW$_a$(He {\sc i} $\lambda$4471) = 0.4\AA) and Fig. \ref{fig3}c -- \ref{fig3}d
(models with EW$_a$(H8 + He {\sc i} $\lambda$3889) = 3.0\AA\ and 
EW$_a$(He {\sc i} $\lambda$4471) = 0.5\AA). The corresponding 
linear regression parameters are given in Table \ref{tab6}.

Our derived values of $Y$ are in general higher than the ones obtained
by \citet{P02} for the same H {\sc ii} regions. The differences come from the
different ways our and their $Y$ values have been derived: 
instead of estimating  
the physical parameters of the H {\sc ii} regions by CLOUDY, 
we have adopted an
approach that does not depend on any particular physical model, external 
estimates or assumptions, i.e. we simply search the multi-parameter space for
the smallest $\chi^2$. Other differences are: 1) we have included H {\sc ii}
regions with higher metallicities to derive a more accurate d$Y$/d$Z$ slope; 
2) \cite{P02} used
expressions for collisional enhancement of He {\sc i} emission lines from 
\citet{B99} while we use the expressions by the same authors \citep{B02} 
but which include not only collisional but also
fluorescent enhancement; 2) \citet{P02} assumed no underlying He {\sc i}
absorption for NGC 346 while we find it to be significant;
3) \citet{P02} calculate their $\chi^2$ in a slightly different manner
from us. They use the expression from \citet{P00} weighting by errors of 
line fluxes, while we use Eq. \ref{eq1} weighting by errors of He abundances.
However, we have checked (for NGC 346) that the choice of a weighting scheme 
makes little difference in the derived He abundance.
The largest differences between \citet{P02}
and our values of $Y$ are for NGC 346 and Mrk 209, our values being 
$\sim$2--3\% higher. 

For the set EW($\lambda$3889) = 3.0\AA\ and EW($\lambda$4471) = 
0.4\AA\ we derive a primordial He mass fraction 
$Y_p$ = 0.2421 $\pm$ 0.0021 from the $Y$ -- O/H linear regression and 
0.2446 $\pm$ 0.0016 from the $Y$ -- N/H linear regression. 
For the set EW($\lambda$3889) = 3.0\AA\ and EW($\lambda$4471) = 
0.5\AA, $Y_p$ = 0.2444 $\pm$ 0.0020 from the 
$Y$ -- O/H linear regression and 
0.2466 $\pm$ 0.0016 from the $Y$ -- N/H linear regression.
The $Y_p$s obtained from the $Y$ -- O/H
linear regressions considering known systematic effects are very similar to 
the value $Y_p$ = 0.243 $\pm$ 0.001 
derived in \S5 for the whole sample without correcting for the difference 
between $T_e$(He {\sc ii}) and $T_e$(O {\sc iii}),
the collisional excitation of hydrogen lines and the 
underlying stellar absorption in the He {\sc i} lines. 
Thus, the combined effect 
of the corrections for all known systematic effects turned out to be very 
small. In the more realistic case where gaseous continuum is taken into 
account, the $Y_p$s differ by only 0.4\%. In the case where it is not taken 
into account, they differ by 0.6\%. 

This means that, even if all known systematic effects are taken into account 
in our previous work on the statistical derivation of $Y_p$ by linear
regressions (ITL94, ITL97 and IT98), the result will not be changed. 
While each individual
systematic effect can change $Y$ upward or downward by as much as 4\%, 
they work in opposite sense and nearly cancel each other out. 
By nearly doubling our sample (from 45 to 82 H {\sc ii} regions) and choosing
the galaxies so that they span as large a metallicity range as possible,
we have considerably reduced the uncertainties in the slopes 
$dY/d$O = 4.3 $\pm$ 0.7 and $dY/dZ$ = 2.8 $\pm$ 0.5 for the whole sample,
as compared to 
$dY/d$O = 5.7 $\pm$ 1.8 and $dY/dZ$ = 3.7 $\pm$ 1.2 for the restricted 
sample with EW$_a$($\lambda$4471) = 
0.4\AA\ and $dY/d$O = 5.1 $\pm$ 1.8 and $dY/dZ$ = 3.4 $\pm$ 1.2 with 
EW$_a$($\lambda$4471) = 
0.5\AA. These values are
consistent within the errors with our previous determinations. 
In a future paper we will apply the $\chi^2$ minimization technique to
correct our whole sample of 82 H {\sc ii} regions for known systematic 
effects and reduce further the uncertainties in $Y_p$
and $dY/dZ$. We will study in detail another
 possible additional systematic effect,
that of the ionization structure of H {\sc ii} regions, which we have not
considered in this paper, simply setting the ionization correction factors
of He {\sc i} to 1.

\section{COSMOLOGICAL IMPLICATIONS}

With our value of $Y_p$ = 0.2421$\pm$0.0021 for EW$_a$($\lambda$4471) = 0.4\AA\
and with
an equivalent number of light neutrino species equal to 3, the SBBN model 
gives $\eta_{10}$ = 10$^{10}$$\eta$ =
3.4$^{+0.7}_{-0.6}$, where $\eta$ is the baryon-to-photon number ratio and the
error bars denote 1$\sigma$ errors. This corresponds to a baryonic mass
fraction $\Omega_b$$h^2$ = 0.012$^{+0.003}_{-0.002}$. 
The respective values of the cosmological parameters with 
$Y_p$ = 0.2444$\pm$0.0020 (obtained for EW$_a$($\lambda$4471) = 0.5\AA)
are $\eta_{10}$ = 10$^{10}$$\eta$ =
4.0$^{+1.1}_{-0.5}$ and $\Omega_b$$h^2$ = 0.015$^{+0.003}_{-0.002}$.
These values are
lower at the 2$\sigma$ level than 
$\Omega_b$$h^2$ = 0.021$\pm$0.002 from recent 
measurements of the deuterium abundance in damped Ly$\alpha$ systems 
\citep{K03}. The latter would correspond to 
$Y_p$ = 0.2476$^{+0.0009}_{-0.0010}$. Recent measurements of the 
fluctuations of the microwave
background radiation by the Wilkinson Microwave Anisotropy Probe (WMAP)
observations have improved considerably the precision of 
$\eta$ \citep{S03}. The WMAP data give 
$\Omega_b$$h^2$ = 0.0224$\pm$0.0009, corresponding to
$Y_p$ = 0.2483$\pm$0.0004 in the SBBN model, significantly 
higher than our values of $Y_p$, and inconsistent at the 95\% C.L.

If, with improved determinations, $\eta$ as obtained 
from the deuterium abundance and microwave background fluctuation
measurements, remains systematically higher than its value obtained from
the He abundance, then this may suggest deviations from SBBN theory.
One possibiblity is that
the equivalent number of the light neutrino species $N_\nu$ is less than 3.
Combining $\Omega_b$$h^2$ = 0.0224 $\pm$ 0.0009 obtained by WMAP \citep{S03}
with $Y_p$ = 0.2421 $\pm$ 0.0021, we obtain $N_\nu$ = 2.48 with
an upper limit $N_\nu$ $<$ 2.83 at the 95\% confidence level \citep{W91}.
With $Y_p$ = 0.2444 $\pm$ 0.0020, we obtain $N_\nu$ = 2.67 with
an upper limit $N_\nu$ $<$ 3.01 at the 95\% confidence level.

\section{SUMMARY AND CONCLUSIONS}

    We present in this paper new high signal-to-noise spectrophotometric
observations of 31 low-metallicity blue compact galaxies (BCGs), containing 33  H {\sc ii} regions and spanning a
large range of heavy element mass fractions from $\sim$ $Z_\odot$/30 to
$Z_\odot$/4. We combine this new data with our previous data \citep{IT98b}
to construct a sample of 82 H {\sc ii} regions and determine the primordial 
helium mass fraction $Y_p$ by linear regressions to the sample. Our sample 
constitutes one of the largest and most homogeneous (obtained, reduced and
analyzed in the same way) data set now available for the determination of 
$Y_p$.
 
We have considered known systematic effects on the He abundance 
determination. For the total sample of 82 H {\sc ii} regions we have 
calculated $N_e$(He {\sc ii}) self-consistently and taken
into account the effects of collisional and fluorescent enhancements. For a
restricted sample of 6 H {\sc ii} regions from our sample and an additional
H {\sc ii} region, NGC 346 in the Small Magellanic Cloud, we have examined,
in addition to the collisional and fluorescent enhancements of He {\sc i} 
emission lines, also the effects of
collisional excitation of hydrogen emission lines, of underlying stellar 
He {\sc i} absorption and of the difference between the temperature 
$T_e$(He {\sc ii}) in the He$^+$ zone and the temperature $T_e$(O {\sc iii})
derived from the [O {\sc iii}]$\lambda$4363/($\lambda$4959+$\lambda$5007)
flux ratio. 
The restricted sample was chosen because the systematic effects on the $Y_p$
determination of 5 of the galaxies in the sample have been discussed
by \citet{P02}, and we can compare our results to theirs.

We have derived the following results:

1. Although each systematic effect may move the helium mass fraction $Y$
up or down by as much as 4\%, the combined result of the systematic effects
on the restricted sample is relatively small ($\la $0.6\%), 
as they act in opposite sense and mostly cancel each other out. 
We derive for the restricted sample $Y_p$ = 0.2421$\pm$0.0021 
adopting EW$_a$($\lambda$4471) = 0.4\AA. This corresponds
to a baryonic mass fraction $\Omega_b$$h^2$ = 0.012$^{+0.003}_{-0.002}$.
If EW$_a$($\lambda$4471) = 0.5\AA\ is adopted then $Y_p$ = 0.2444$\pm$0.0020
corresponding to $\Omega_b$$h^2$ = 0.015$^{+0.003}_{-0.002}$. These values
of $\Omega_b$$h^2$ are
lower than the values derived from the deuterium abundance 
and microwave background radiation fluctuation measurements. 
This may indicate that the equivalent number of light neutrino 
species $N_\nu$ is less than 3 and hence that there are deviations from the 
standard Big Bang nucleosynthesis model. 

2. The slopes $dY/d$O and $dY/dZ$ derived from the $Y$ -- O/H linear 
regressions for the restricted sample with two adopted values 
of EW$_a$($\lambda$4471) = 0.4\AA\ and 0.5\AA\ are respectively 
5.7 $\pm$ 1.8,
3.7 $\pm$ 1.2, and 5.1 $\pm$ 1.8,
3.4 $\pm$ 1.2. They are consistent with previous determinations by 
\citet{ITL97} and 
\citet{IT98b} using BCGs, and by \citet{J03} from nearby K dwarf stars.

     3. We have considerably reduced the errors in the $dY/d$O and
$dY/dZ$ slopes derived for the whole sample as it contains galaxies 
spanning a wide range
of metallicities, which was not the case in our previous work.
From the $Y$ -- O/H linear regression of the whole sample, with only 
collisional and fluorescent enhancements taken into account, we derive 
slopes $dY/d$O = 4.3 $\pm$ 0.7 and $dY/dZ$ = 2.8 $\pm$ 0.5, 
in good agreement with the slopes derived
from the subsample of 7 H {\sc ii} regions where all known systematic effects,
with the exception of ionization correction effects,
are taken into account. 

\acknowledgements
Y.I.I. is grateful to the staff of the Astronomy Department at the 
University of Virginia for warm hospitality. 
Y.I.I. and T.X.T. thank the support of National Science Foundation
grant AST 02-05785.

\clearpage

\begin{deluxetable}{lccccclc}
\tabletypesize{\footnotesize}
\tablenum{1}
\tablecolumns{8}
\tablewidth{510pt}
\tablecaption{General Parameters of Galaxies}
\tablehead{
\colhead{} & \multicolumn{2}{c}{Coordinates (2000.0)} &\colhead{} 
&\multicolumn{2}{c}{ } &\colhead{} \\ \\ \cline{2-3} \\ 
\colhead{Name}&\colhead{$\alpha$}&\colhead{$\delta$}&\colhead{$m_{pg}$}&
\colhead{$z$\tablenotemark{a}}
&\colhead{$M_{pg}$\tablenotemark{b}} &\colhead{Other names}
&\colhead{Ref.}  }
\startdata 
HS $2359+1659$ & 00$^h$02$^m$09\fs9~     &\,\,\,$+17$\arcdeg15\arcmin59\arcsec&16.6   &0.02076 $\pm$ 0.00012 &--18.0&        &1 \\
UM 238         & 00\ \ 24\ \ 41.9~        &$+01$\ 44\ 03                      &16.5   &0.01423 $\pm$ 0.00017 &--17.3&        &2 \\
HS $0029+1748$ & 00\ \ 32\ \ 03.2~        &$+18$\ 04\ 44                      &17.6   &0.00705 $\pm$ 0.00011 &--14.7&        &1 \\
HS $0111+2115$ & 01\ \ 14\ \ 37.6~        &$+21$\ 31\ 16                      &16.1   &0.03227 $\pm$ 0.00010 &--19.5&        &1 \\
HS $0122+0743$ & 01\ \ 25\ \ 34.2~        &$+07$\ 59\ 22                      &15.7   &0.00986 $\pm$ 0.00006 &--17.3&UGC 993 &1 \\
HS $0128+2832$ & 01\ \ 31\ \ 21.3~        &$+28$\ 48\ 12                      &17.6   &0.01613 $\pm$ 0.00010 &--16.4&        &1 \\
HS $0134+3415$ & 01\ \ 37\ \ 13.8~        &$+34$\ 31\ 12                      &18.0   &0.01949 $\pm$ 0.00007 &--16.5&        &1 \\
UM 133         & 01\ \ 44\ \ 41.4~        &$+04$\ 53\ 27                      &15.7   &0.00532 $\pm$ 0.00004 &--15.9&        &3 \\
UM 396         & 02\ \ 07\ \ 26.5~        &$+02$\ 56\ 55                      &\nodata&0.02093 $\pm$ 0.00011 &\nodata&       &2 \\
Mrk 1063       & 02\ \ 54\ \ 33.6~        &$-10$\ 01\ 40                      &13.6   &0.00503 $\pm$ 0.00007 &--17.9&NGC 1140, VV 482& 7 \\
J $0519+0007$  & 05\ \ 19\ \ 02.7~        &$+00$\ 07\ 29                      &18.2   &0.04476 $\pm$ 0.00006 &--18.1&        &16 \\
HS $0735+3512$ & 07\ \ 38\ \ 58.4~        &$+35$\ 05\ 36                      &17.6   &0.03020 $\pm$ 0.00003 &--17.8&        &4\\
HS $0811+4913$ & 08\ \ 14\ \ 47.0~        &$+49$\ 04\ 10                      &18.6   &0.00177 $\pm$ 0.00009 &--10.7&        &4 \\
HS $0837+4717$ & 08\ \ 40\ \ 29.9~        &$+47$\ 07\ 09                      &18.2   &0.04195 $\pm$ 0.00004 &--17.9&        &4,5 \\
HS $0924+3821$ & 09\ \ 28\ \ 06.3~        &$+38$\ 07\ 55                      &18.5   &0.06088 $\pm$ 0.00006 &--18.4&        &4 \\
CGCG 007$-$025 & 09\ \ 44\ \ 01.9~        &$-00$\ 38\ 32                      &17.5   &0.00483 $\pm$ 0.00004 &--13.9&        &6 \\
Mrk 1236       & 09\ \ 49\ \ 54.1~        &$+00$\ 36\ 59                      &13.5   &0.00625 $\pm$ 0.00010 &--18.5&VV 620  &7 \\
HS $1028+3843$ & 10\ \ 31\ \ 51.8~        &$+38$\ 28\ 07                      &19.4   &0.02945 $\pm$ 0.00005 &--16.0&        &8 \\
Mrk 724        & 10\ \ 41\ \ 09.6~        &$+21$\ 21\ 43                      &16.5   &0.00402 $\pm$ 0.00005 &--14.5&        &9 \\
Mrk 35         & 10\ \ 45\ \ 22.4~        &$+55$\ 57\ 37                      &13.3   &0.00323 $\pm$ 0.00008 &--17.3&NGC 3353, Haro 3  & 15 \\
UM 422         & 11\ \ 20\ \ 14.6~        &$+02$\ 31\ 51                      &\nodata&0.00536 $\pm$ 0.00002 &\nodata&        &2 \\
UM 439         & 11\ \ 36\ \ 36.8~        &$+00$\ 48\ 58                      &15.1   &0.00359 $\pm$ 0.00004 &--15.7&UGC 6578   &2 \\
POX 36         & 11\ \ 58\ \ 58.3~        &$-19$\ 01\ 41                      &14.3   &0.00355 $\pm$ 0.00007 &--16.5&I SZ 63        &10 \\
Mrk 1315       & 12\ \ 15\ \ 18.7~        &$+20$\ 38\ 28                      &16.5   &0.00261 $\pm$ 0.00005 &--13.6&        &2 \\
HS $1213+3636$A& 12\ \ 15\ \ 34.4~        &$+36$\ 20\ 16                      &17.5   &0.00092 $\pm$ 0.00004 &--10.3&        &11 \\
HS $1214+3801$ & 12\ \ 17\ \ 09.7~        &$+37$\ 44\ 52                      &16.5   &0.00110 $\pm$ 0.00004 &--11.7&        &12 \\
Mrk 1329       & 12\ \ 37\ \ 03.0~        &$+06$\ 55\ 36                      &14.6   &0.00519 $\pm$ 0.00012 &--17.0&        &7 \\
HS $1311+3628$ & 13\ \ 13\ \ 18.5~        &$+36$\ 12\ 09                      &18.0   &0.00293 $\pm$ 0.00006 &--12.3&        &11 \\
Mrk 450        & 13\ \ 14\ \ 47.2~        &$+34$\ 53\ 01                      &16.0   &0.00279 $\pm$ 0.00005 &--14.2&UGC 8323, VV 616 &13 \\
Mrk 67         & 13\ \ 41\ \ 55.9~        &$+30$\ 30\ 29                      &\nodata&0.00311 $\pm$ 0.00002 &\nodata&UGCA 372        &14 \\
HS $2236+1344$ & 22\ \ 38\ \ 31.1~        &$+14$\ 00\ 29                      &18.2   &0.02115 $\pm$ 0.00034 &--16.4&        &1 \\
\enddata 
\tablenotetext{a}{derived from spectra in this paper.}
\tablenotetext{b}{all objects are assumed to be at their redshift 
distances with $H_0$=75 km\ s$^{-1}$Mpc$^{-1}$.}
\tablerefs{(1) \citet{U03}; (2) \citet{T91}; 
(3) \citet{K01b};
(4) \citet{P99}; (5) \citet{K00}; (6) \citet{van00}; 
(7) \citet{G00};
(8) \citet{K01a}; (9) \citet{KS86}; (10) \citet{K81}; (11) \citet{U01}; 
(12) \citet{H00};
(13) \citet{KD81}; (14) \citet{F80}; (15) \citet{St96}; (16) \citet{Kn03}.}

\end{deluxetable}

\clearpage

\begin{deluxetable}{llccccc}
\tabletypesize{\footnotesize}
\tablenum{2}
\tablecolumns{6}
\tablewidth{0pt}
\tablecaption{Journal of Observations}
\tablehead{
\colhead{Galaxy}&\colhead{Date}&\colhead{Number of}&\colhead{Exposure time}&\colhead{Airmass} 
& \colhead{Position angle} & \colhead{Aperture\tablenotemark{a}} \\ 
\colhead{}&\colhead{}&\colhead{Exposures}&\colhead{(minutes)}
&\colhead{}&\colhead{(degrees)}&\colhead{(arcsec)} }
\startdata
HS $2359+1659$ & December 31,2002 &     3      &     45        &  1.1 & 158~ & 2.0 $\times$ 6.9      \\
UM 238         & December 31      &     3      &     45        &  1.3 & 100~ & 2.0 $\times$ 4.5      \\
HS $0029+1748$ & January 1,2003   &     3      &     45        &  1.2 & 180~ & 2.0 $\times$ 6.9      \\
HS $0111+2115$ & December 30      &     2      &     30        &  1.1 &   0  & 2.0 $\times$ 4.1      \\
HS $0122+0743$ & December 31      &     3      &     45        &  1.2 &   0  & 2.0 $\times$ 4.8      \\
HS $0128+2832$ & December 31      &     3      &     45        &  1.3 &   0  & 2.0 $\times$ 4.1      \\
HS $0134+3415$ & January 1        &     3      &     45        &  1.4 & 100~ & 2.0 $\times$ 6.9      \\
UM 133         & December 30      &     4      &     60        &  1.2 &  22  & 2.0 $\times$ 6.9      \\
UM 396         & December 31      &     2      &     30        &  1.3 &   0  & 2.0 $\times$ 6.9      \\
Mrk 1063       & January 1        &     3      &     30        &  1.5 & 180~ & 2.0 $\times$11.0      \\
J $0519+0007$  & December 30      &     3      &     60        &  1.2 &  22  & 2.0 $\times$ 6.9      \\
HS $0735+3512$ & January 1        &     3      &     45        &  1.1 & 100  & 2.0 $\times$ 5.5      \\
HS $0811+4913$ & December 31      &     3      &     60        &  1.2 &   0  & 2.0 $\times$ 6.9      \\
HS $0837+4717$ & January 1        &     4      &     60        &  1.1 & 100~ & 2.0 $\times$ 6.9      \\
HS $0924+3821$ & January 1        &     3      &     45        &  1.0 & 100~ & 2.0 $\times$ 6.9      \\
CGCG $007-025$\tablenotemark{b} & December 30      &     3      &     60        &  1.4 & 155~ & 2.0 $\times$ 3.5      \\
Mrk 1236       & December 31      &     3      &     45        &  1.2 &  80  & 2.0 $\times$ 4.1      \\
HS $1028+3843$ & December 31      &     3      &     60        &  1.3 &   0  & 2.0 $\times$ 4.1      \\
Mrk 724        & December 30      &     3      &     45        &  1.1 & 155~ & 2.0 $\times$ 6.9      \\
Mrk 35         & December 30      &     3      &     30        &  1.5 &  45  & 2.0 $\times$ 6.9      \\
UM 422         & December 31      &     3      &     45        &  1.2 & 170~ & 2.0 $\times$ 5.5      \\
UM 439         & December 31      &     2      &     30        &  1.2 & 170~ & 2.0 $\times$ 5.5      \\
POX 36         & December 30      &     3      &     30        &  1.7 & 155~ & 2.0 $\times$ 5.5      \\
Mrk 1315       & December 30      &     3      &     45        &  1.2 & 155~ & 2.0 $\times$ 6.9      \\
HS $1213+3636$A& January 1        &     3      &     45        &  1.3 &  45  & 2.0 $\times$ 6.9      \\
HS $1214+3801$ & December 31      &     3      &     45        &  1.2 & 135~ & 2.0 $\times$ 6.9      \\
Mrk 1329       & December 30      &     3      &     45        &  1.1 &  60  & 2.0 $\times$ 4.1      \\
HS $1311+3628$ & January 1        &     3      &     45        &  1.2 &  45  & 2.0 $\times$ 6.9      \\
Mrk 450\tablenotemark{b}        & January 1        &     2      &     30        &  1.1 &  45  & 2.0 $\times$ 5.5      \\
Mrk 67         & January 1        &     2      &     20        &  1.1 &  45  & 2.0 $\times$ 6.9      \\
HS $2236+1344$ & December 30      &     2      &     40        &  1.3 &   0  & 2.0 $\times$ 9.7      \\
\enddata
\tablenotetext{a}{Aperture used for extraction of one-dimensional spectra.}
\tablenotetext{b}{Galaxies in which two H {\sc ii} regions are used for 
abundance determination. The two spectra are extracted within equal apertures.}
\end{deluxetable}


\clearpage


\clearpage

\begin{figure*}
\figurenum{1}
\epsscale{1.1}
\vspace*{-1.5cm}\hspace*{-2.0cm}\psfig{figure=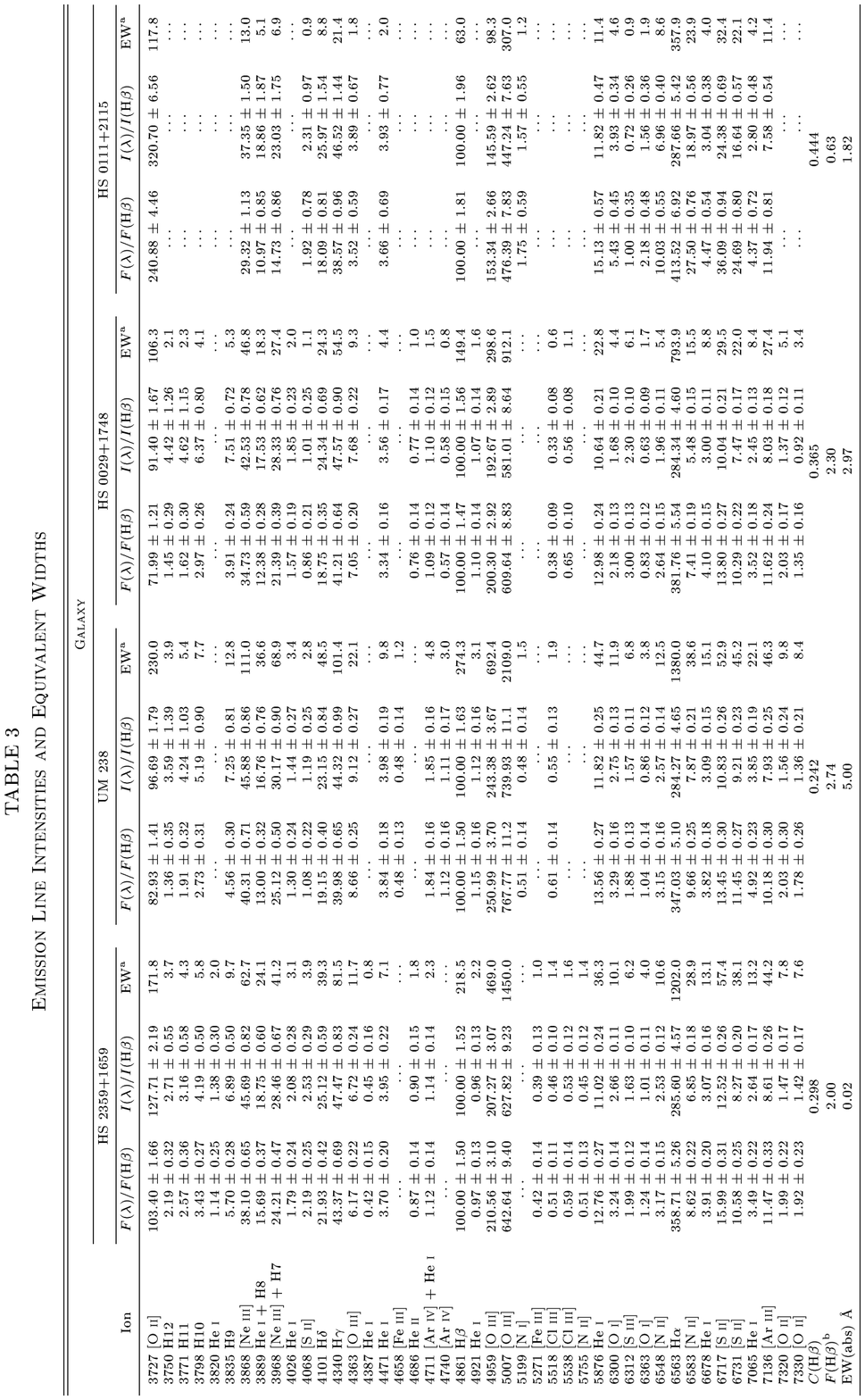,angle=180,height=25cm}
\end{figure*}

\clearpage

\begin{figure*}
\figurenum{}
\epsscale{1.1}
\vspace*{-1.5cm}\hspace*{-2.0cm}\psfig{figure=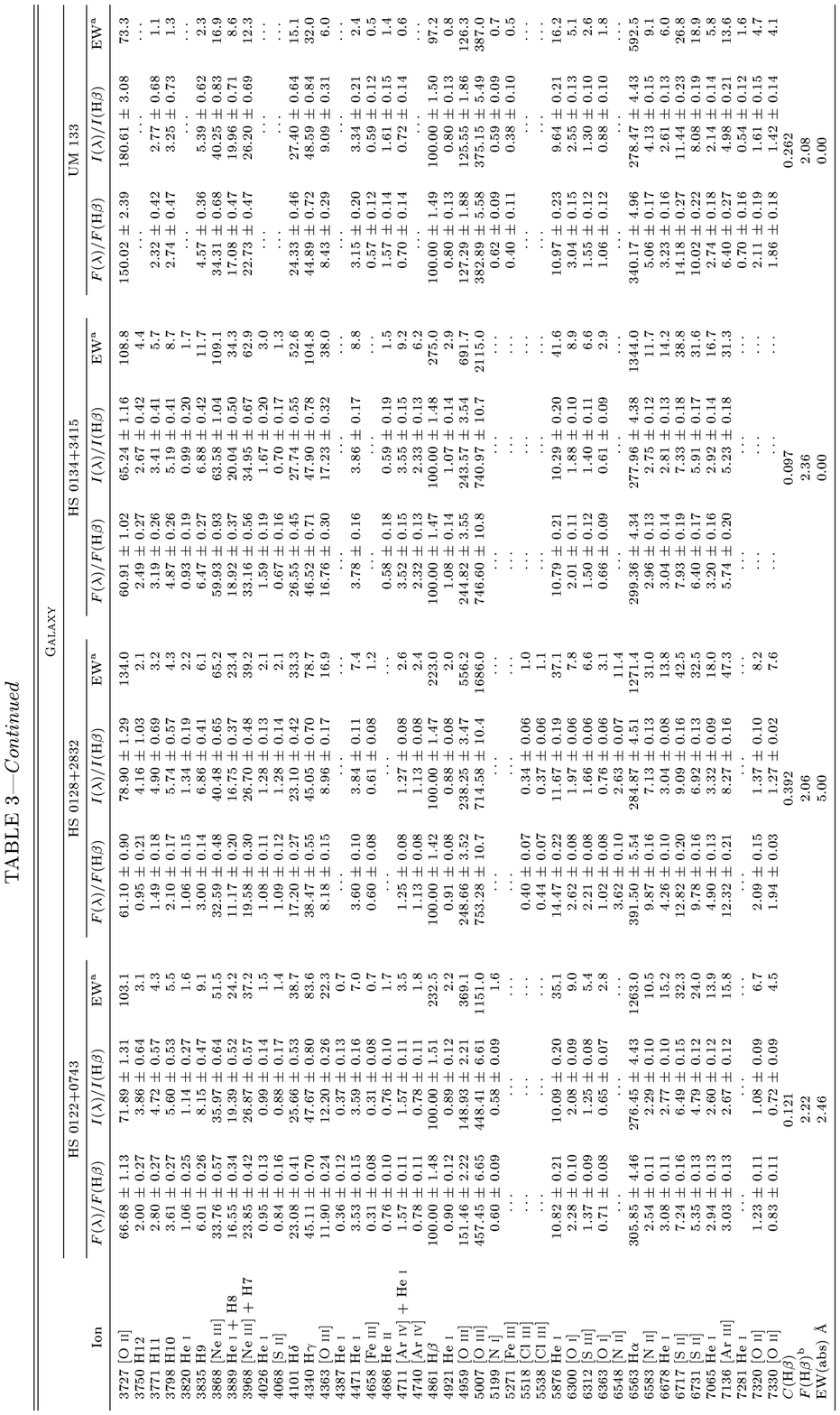,angle=180,height=25cm}
\end{figure*}

\clearpage

\begin{figure*}
\figurenum{}
\epsscale{1.1}
\vspace*{-1.5cm}\hspace*{-2.0cm}\psfig{figure=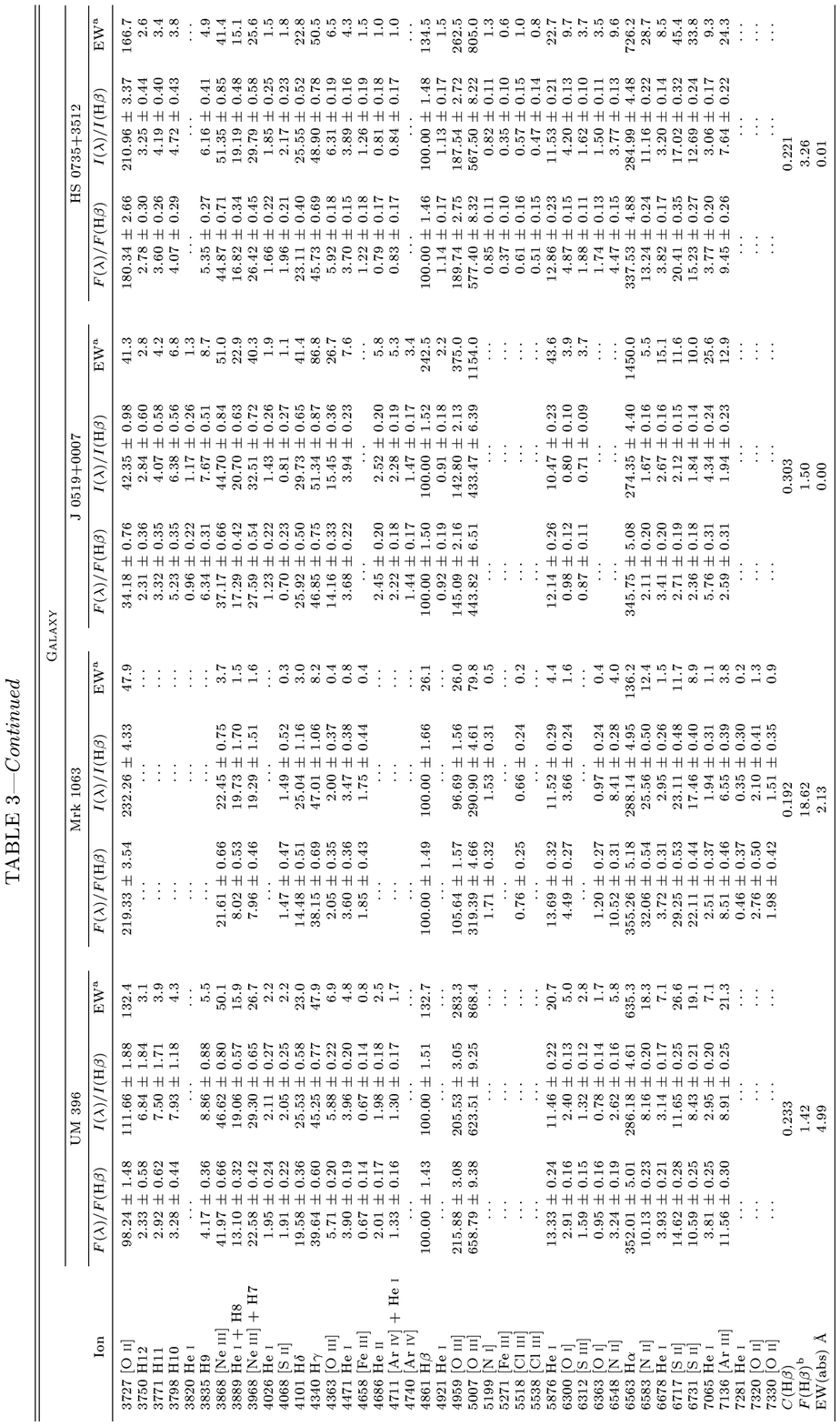,angle=180,height=25cm}
\end{figure*}

\clearpage

\begin{figure*}
\figurenum{}
\epsscale{1.1}
\vspace*{-1.5cm}\hspace*{-2.0cm}\psfig{figure=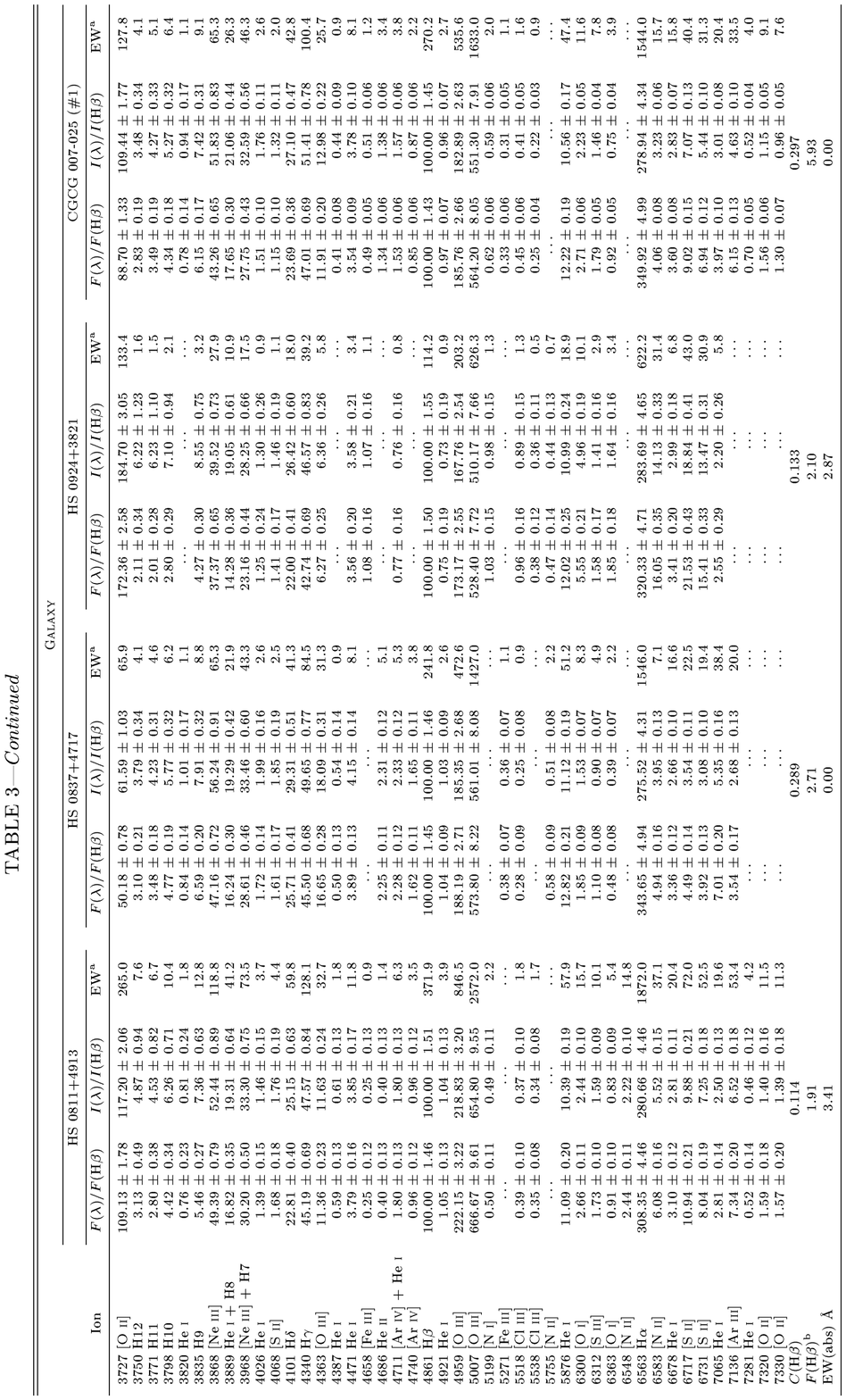,angle=180,height=25cm}
\end{figure*}

\clearpage

\begin{figure*}
\figurenum{}
\epsscale{1.1}
\vspace*{-1.5cm}\hspace*{-2.0cm}\psfig{figure=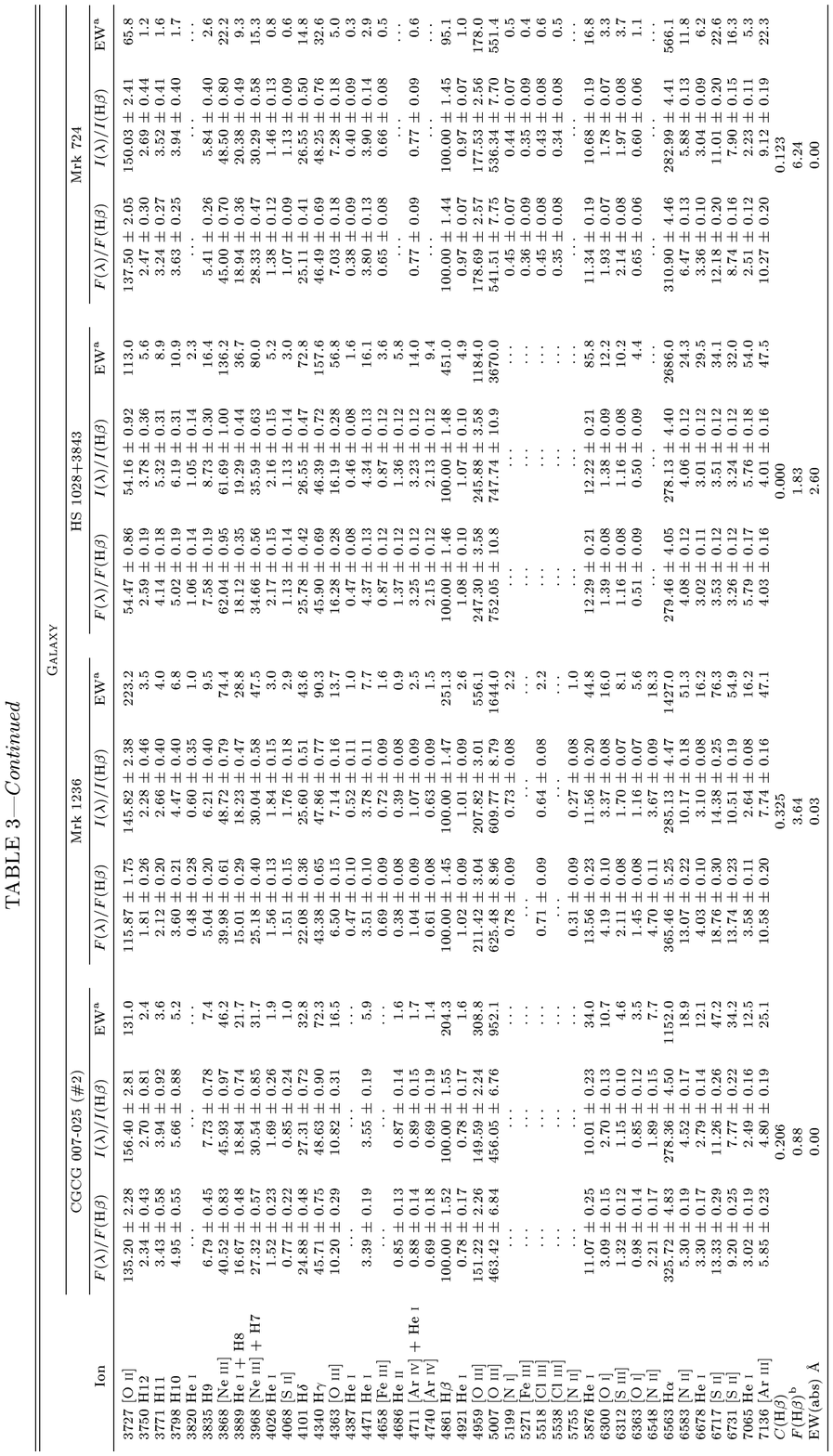,angle=180,height=25cm}
\end{figure*}

\clearpage

\begin{figure*}
\figurenum{}
\epsscale{1.1}
\vspace*{-1.5cm}\hspace*{-2.0cm}\psfig{figure=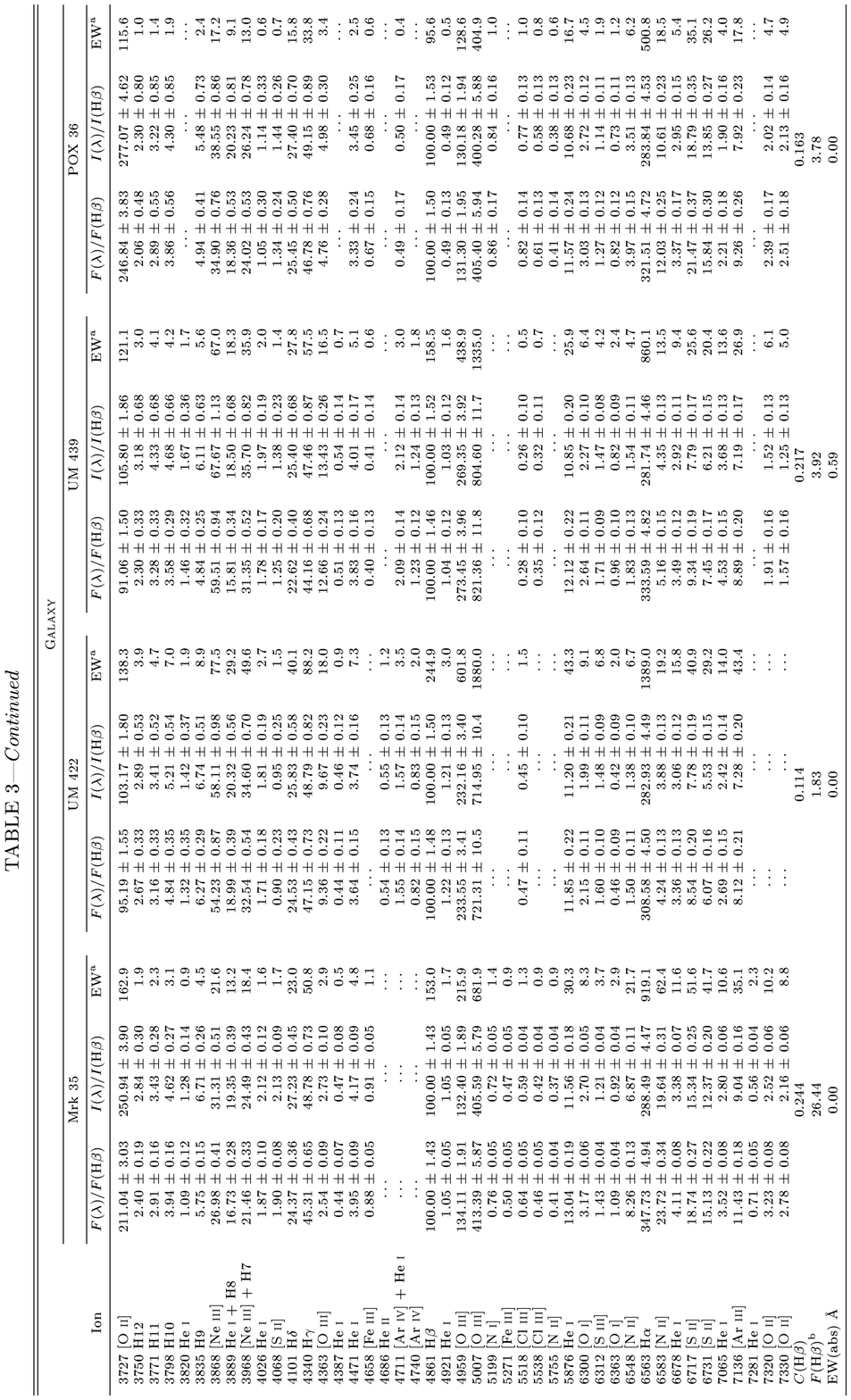,angle=180,height=25cm}
\end{figure*}

\clearpage

\begin{figure*}
\figurenum{}
\epsscale{1.1}
\vspace*{-1.5cm}\hspace*{-2.0cm}\psfig{figure=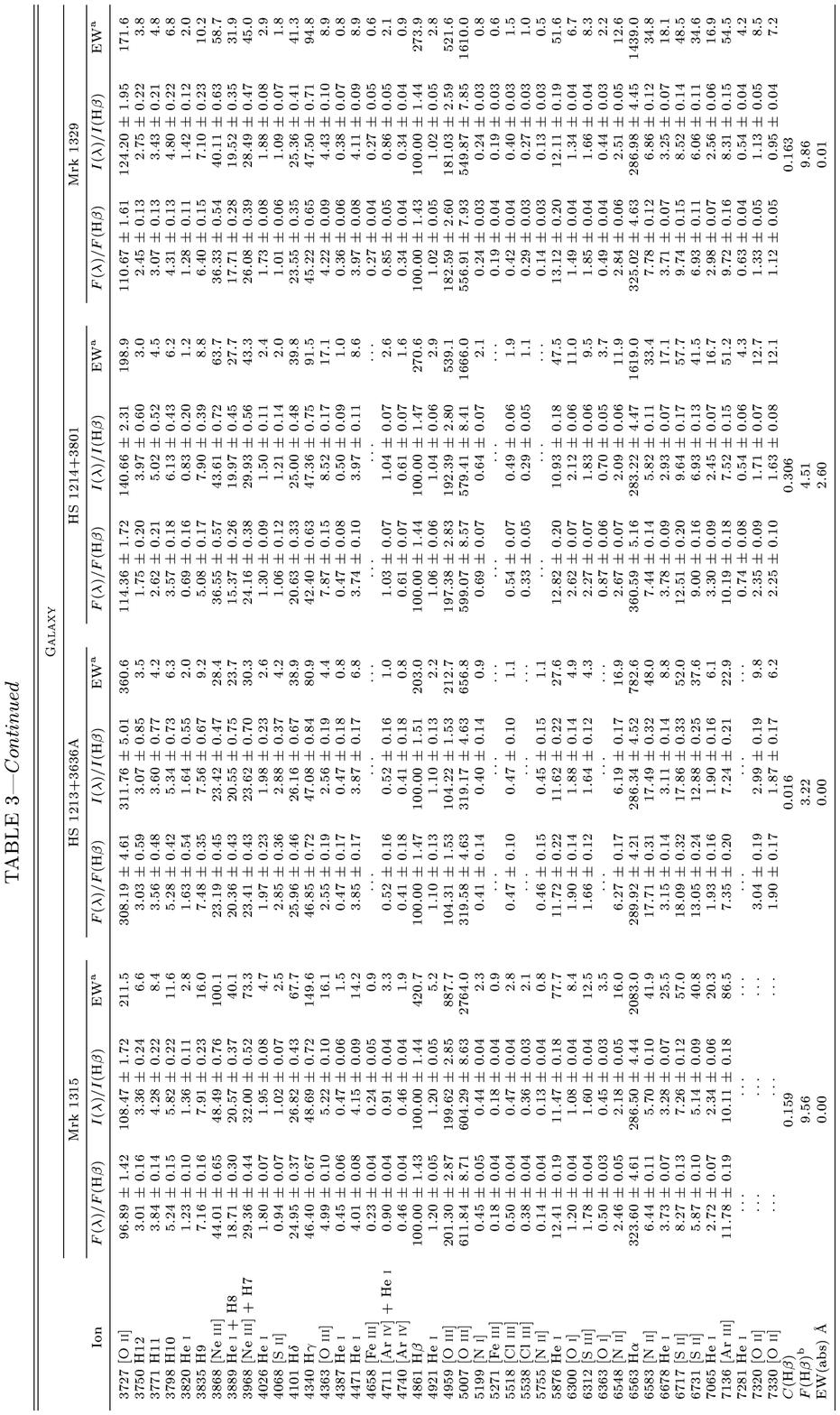,angle=180,height=25cm}
\end{figure*}

\clearpage

\begin{figure*}
\figurenum{}
\epsscale{1.1}
\vspace*{-1.5cm}\hspace*{-2.0cm}\psfig{figure=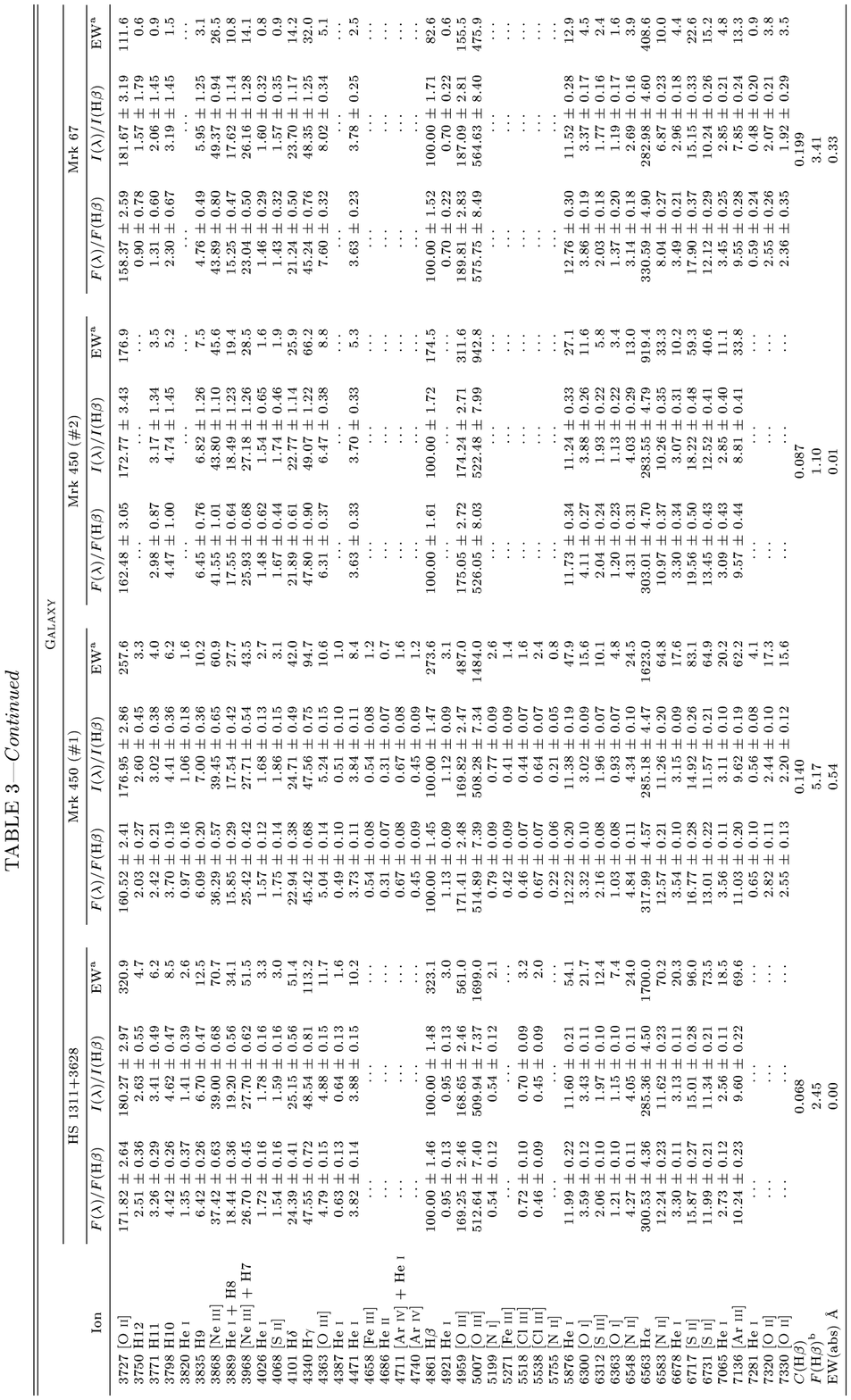,angle=180,height=25cm}
\end{figure*}

\clearpage

\begin{figure*}
\figurenum{}
\epsscale{1.1}
\vspace*{-1.5cm}\hspace*{-2.0cm}\psfig{figure=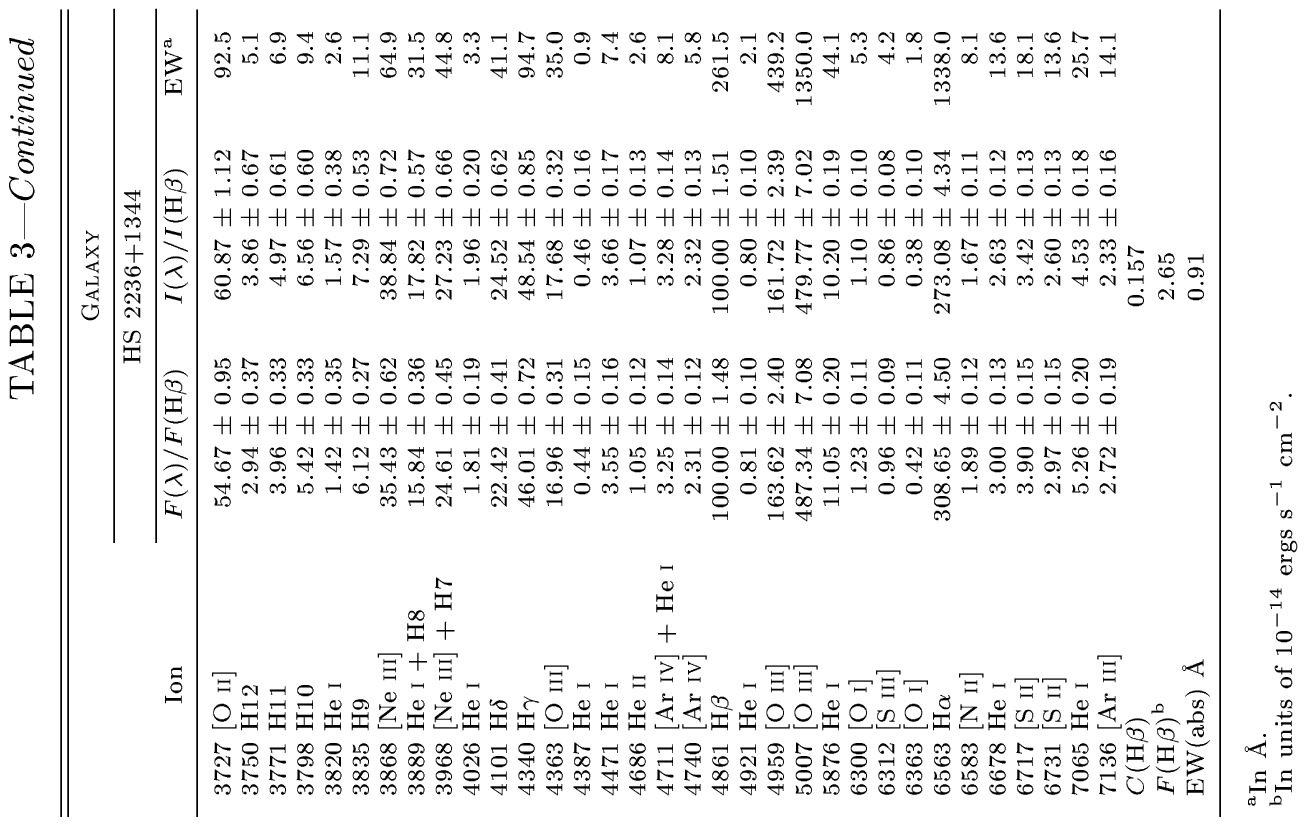,angle=180,height=25cm}
\end{figure*}

\clearpage

\begin{figure*}
\figurenum{}
\epsscale{1.1}
\vspace*{-4.5cm}\hspace*{-4.0cm}\psfig{figure=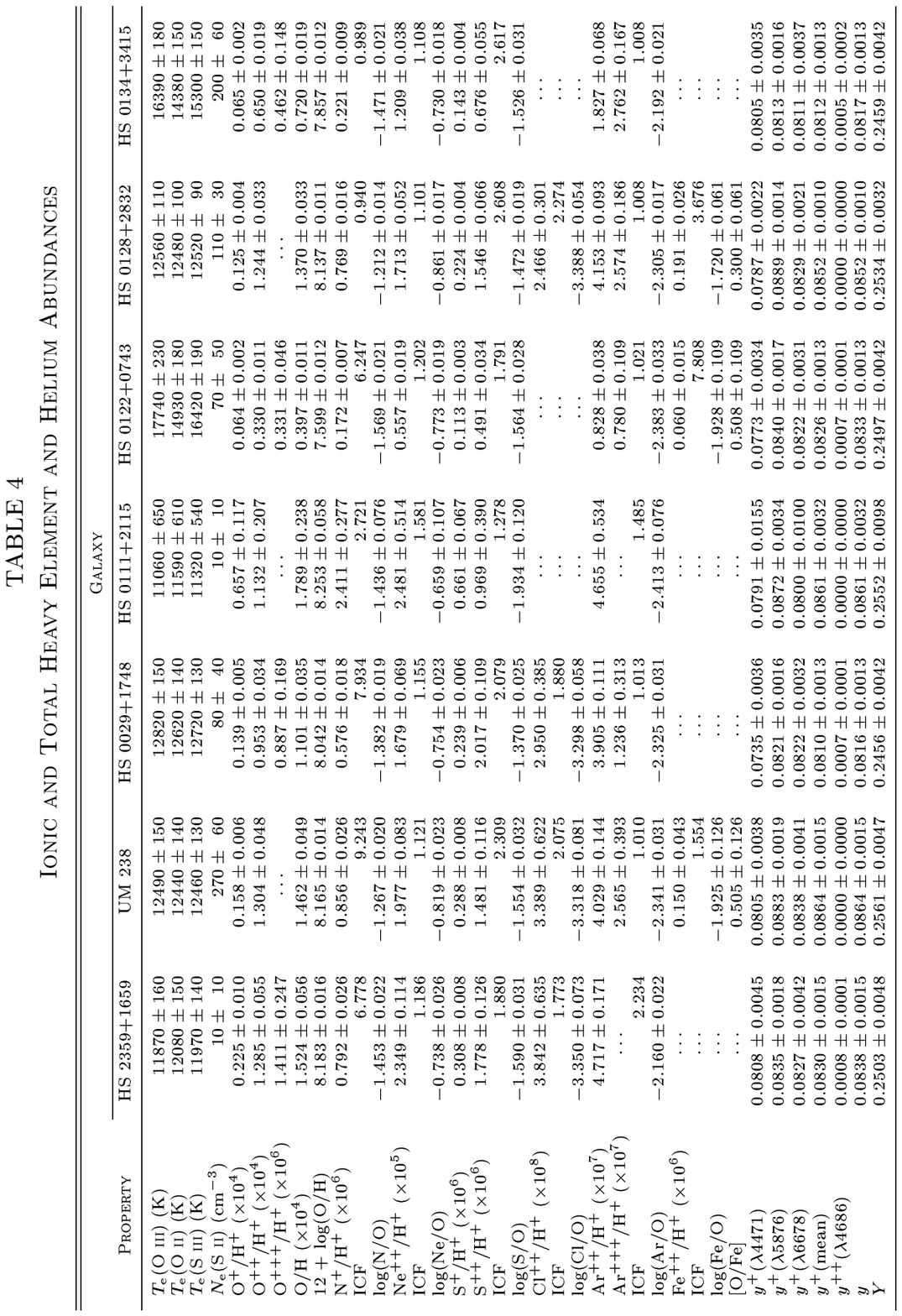,angle=180,height=30cm}
\end{figure*}

\clearpage

\begin{figure*}
\figurenum{}
\epsscale{1.2}
\vspace*{-4.5cm}\hspace*{-4.0cm}\psfig{figure=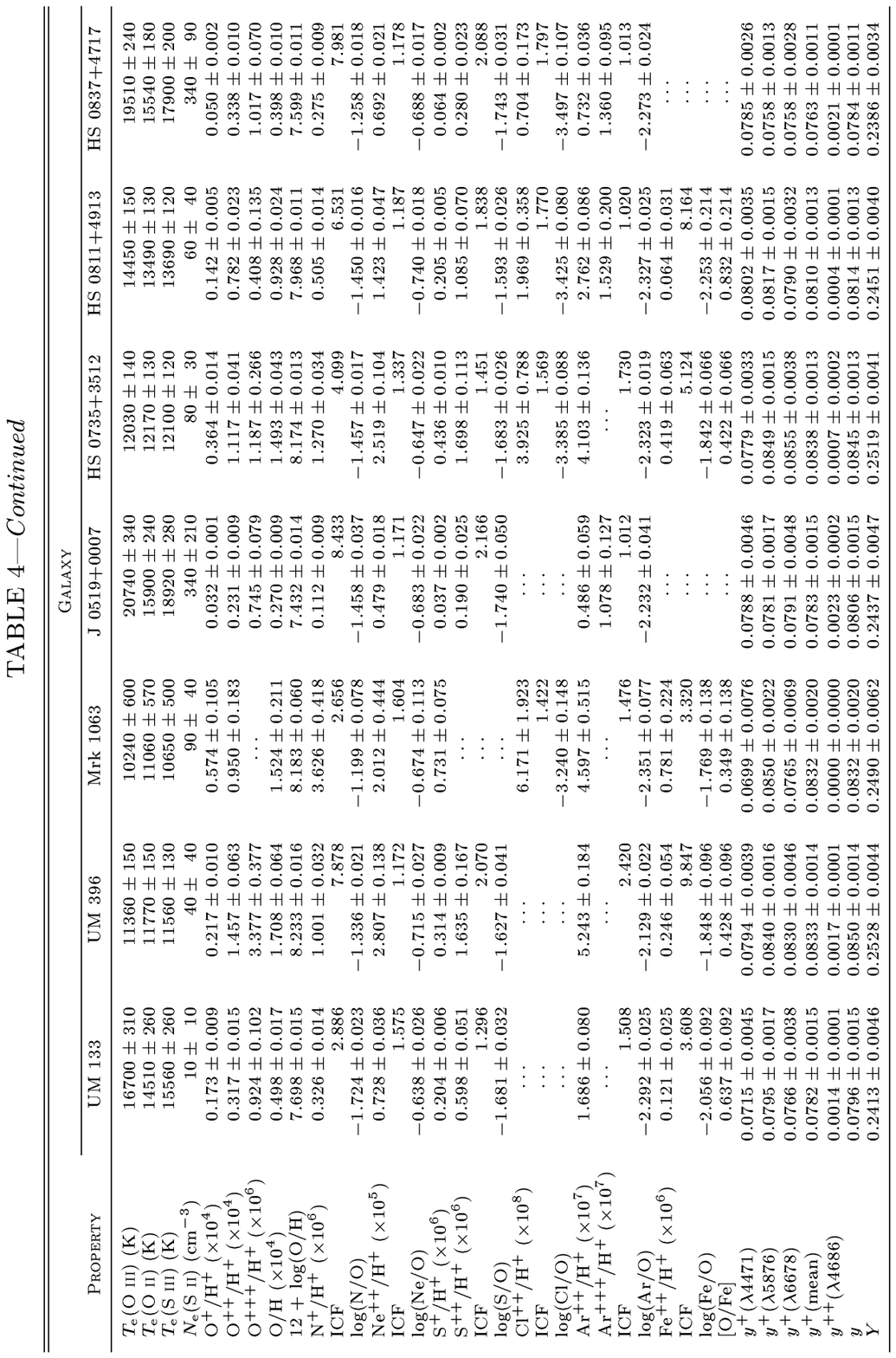,angle=180,height=30cm}
\end{figure*}

\clearpage

\begin{figure*}
\figurenum{}
\epsscale{1.2}
\vspace*{-4.5cm}\hspace*{-4.0cm}\psfig{figure=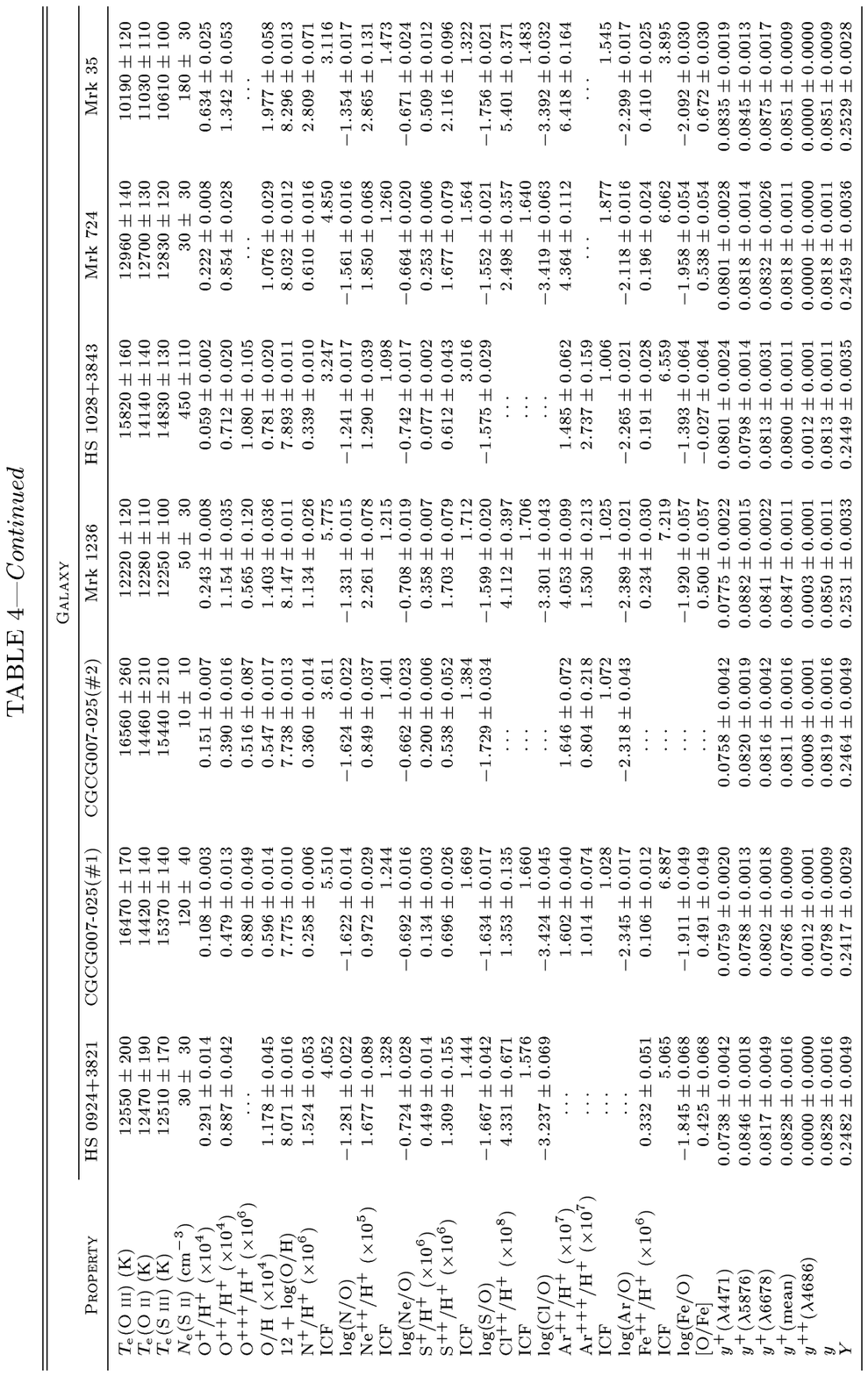,angle=180,height=30cm}
\end{figure*}

\clearpage

\begin{figure*}
\figurenum{}
\epsscale{1.2}
\vspace*{-4.5cm}\hspace*{-4.0cm}\psfig{figure=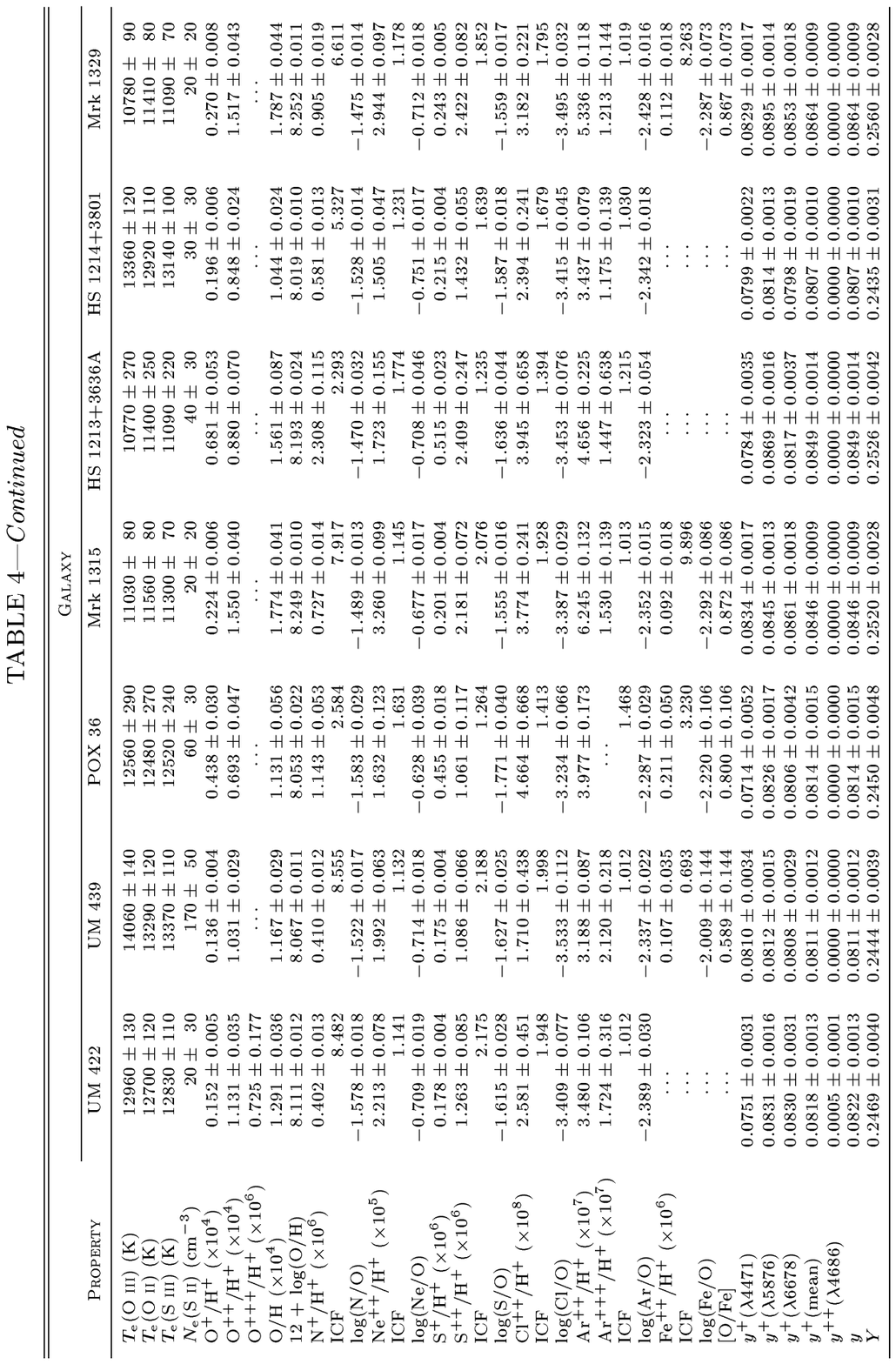,angle=180,height=30cm}
\end{figure*}

\clearpage

\begin{figure*}
\figurenum{}
\epsscale{1.2}
\vspace*{-4.5cm}\hspace*{-4.0cm}\psfig{figure=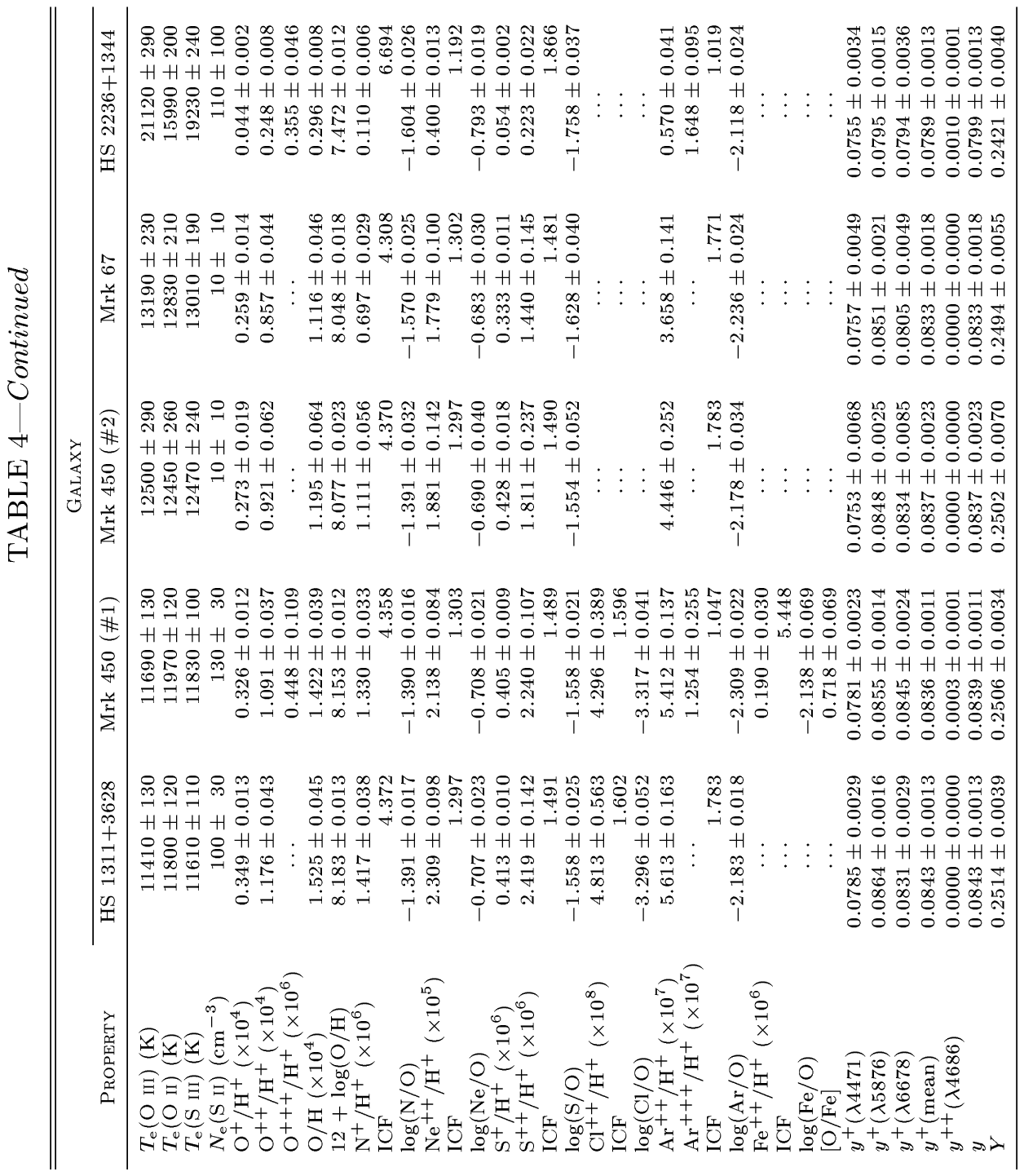,angle=180,height=30cm}
\end{figure*}


\clearpage

  \begin{deluxetable}{lrrrr}
  \tablenum{5}
  \tablecolumns{5}
  \tablewidth{0pc}
  \tabletypesize{\footnotesize}
  \tablecaption{Sample Used for Primordial Helium Abundance Determination}
  \tablehead{
  \colhead{Object} & \colhead{O/H $\times$10$^4$} &
  \colhead{N/H $\times$10$^6$} & \colhead{$Y$} &
  \colhead{References}
  }
  \startdata
I Zw 18 SE        &  1.53 $\pm$   0.06 &   4.06 $\pm$    0.51 & 0.2414 $\pm$  0.0063 & 2   \\
I Zw 18 SE        &  1.55 $\pm$   0.10 &   4.06 $\pm$    0.51 & 0.2389 $\pm$  0.0057 & 3   \\
SBS 0335-052E     &  1.99 $\pm$   0.08 &   5.23 $\pm$    0.81 & 0.2515 $\pm$  0.0052 & 4   \\
SBS 0335-052E     &  1.99 $\pm$   0.05 &   5.23 $\pm$    0.81 & 0.2466 $\pm$  0.0029 & 2   \\
SBS 0335-052E     &  2.02 $\pm$   0.05 &   5.23 $\pm$    0.81 & 0.2475 $\pm$  0.0027 & 5   \\
J 0519+0007       &  2.70 $\pm$   0.09 &   9.42 $\pm$    0.74 & 0.2437 $\pm$  0.0047 & 1   \\
SBS 0940+544      &  2.71 $\pm$   0.11 &   6.61 $\pm$    0.40 & 0.2455 $\pm$  0.0066 & 6   \\
SBS 0940+544      &  2.74 $\pm$   0.12 &   6.92 $\pm$    0.71 & 0.2431 $\pm$  0.0070 & 7   \\
HS 2236+1344      &  2.96 $\pm$   0.08 &   7.37 $\pm$    0.39 & 0.2421 $\pm$  0.0040 & 1   \\
SBS 0940+544      &  3.14 $\pm$   0.08 &   7.05 $\pm$    0.28 & 0.2468 $\pm$  0.0034 & 7   \\
SBS 1159+545      &  3.14 $\pm$   0.10 &   8.20 $\pm$    0.37 & 0.2409 $\pm$  0.0043 & 6   \\
Tol 1214--277     &  3.45 $\pm$   0.10 &   7.90 $\pm$    0.50 & 0.2432 $\pm$  0.0038 & 8   \\
UGC 4483          &  3.45 $\pm$   0.09 &   7.73 $\pm$    0.24 & 0.2439 $\pm$  0.0037 & 6   \\
Tol 65            &  3.48 $\pm$   0.10 &   7.92 $\pm$    0.37 & 0.2495 $\pm$  0.0043 & 8   \\
SBS 1415+437 (\#1)&  3.88 $\pm$   0.09 &   8.50 $\pm$    0.25 & 0.2451 $\pm$  0.0032 & 6   \\
HS 0122+0743      &  3.97 $\pm$   0.11 &  10.72 $\pm$    0.42 & 0.2497 $\pm$  0.0042 & 1   \\
SBS 1415+437 (\#1)&  3.97 $\pm$   0.10 &  11.41 $\pm$    0.36 & 0.2451 $\pm$  0.0038 & 9   \\
HS 0837+4717      &  3.98 $\pm$   0.10 &  21.97 $\pm$    0.69 & 0.2386 $\pm$  0.0034 & 1   \\
SBS 1415+437 (\#1)&  4.10 $\pm$   0.07 &  11.29 $\pm$    1.20 & 0.2460 $\pm$  0.0030 & 10  \\
SBS 1415+437 (\#2)&  4.12 $\pm$   0.28 &  12.64 $\pm$    2.00 & 0.2430 $\pm$  0.0100 & 10  \\
HS 1442+4250      &  4.31 $\pm$   0.19 &  15.63 $\pm$    1.17 & 0.2432 $\pm$  0.0075 & 11  \\
SBS 1211+540      &  4.36 $\pm$   0.11 &   9.79 $\pm$    0.36 & 0.2510 $\pm$  0.0038 & 6   \\
VII Zw 403        &  4.93 $\pm$   0.19 &  14.63 $\pm$    0.62 & 0.2415 $\pm$  0.0052 & 6   \\
UM 133            &  4.98 $\pm$   0.17 &   9.41 $\pm$    0.39 & 0.2413 $\pm$  0.0046 & 1   \\
SBS 1249+493      &  5.38 $\pm$   0.18 &  13.65 $\pm$    0.60 & 0.2452 $\pm$  0.0055 & 6   \\
CGCG 007-025 (\#2)&  5.47 $\pm$   0.17 &  13.00 $\pm$    0.50 & 0.2464 $\pm$  0.0049 & 1   \\
SBS 1331+493      &  5.60 $\pm$   0.16 &  15.85 $\pm$    0.60 & 0.2507 $\pm$  0.0047 & 6   \\
SBS 1128+573      &  5.67 $\pm$   0.45 &  17.54 $\pm$    2.67 & 0.2425 $\pm$  0.0169 & 6   \\
SBS 1420+544      &  5.70 $\pm$   0.16 &  15.81 $\pm$    0.71 & 0.2447 $\pm$  0.0038 & 6   \\
SBS 1205+557      &  5.75 $\pm$   0.43 &  18.12 $\pm$    1.57 & 0.2437 $\pm$  0.0128 & 6   \\
Mrk 209           &  5.94 $\pm$   0.14 &  19.33 $\pm$    0.44 & 0.2472 $\pm$  0.0027 & 6   \\
CGCG 007-025 (\#1)&  5.96 $\pm$   0.14 &  14.24 $\pm$    0.33 & 0.2417 $\pm$  0.0029 & 1   \\
UM 461            &  6.09 $\pm$   0.41 &  19.31 $\pm$    2.80 & 0.2466 $\pm$  0.0115 & 4   \\
SBS 1030+583      &  6.29 $\pm$   0.20 &  15.75 $\pm$    0.69 & 0.2444 $\pm$  0.0051 & 6   \\
Mrk 71 (\#2)      &  6.62 $\pm$   0.30 &  17.96 $\pm$    1.04 & 0.2383 $\pm$  0.0071 & 6   \\
Mrk 600           &  6.70 $\pm$   0.24 &  14.42 $\pm$    0.57 & 0.2397 $\pm$  0.0057 & 4   \\
Mrk 36            &  6.70 $\pm$   0.35 &  21.68 $\pm$    1.46 & 0.2395 $\pm$  0.0089 & 4   \\
Mrk 71 (\#1)      &  7.12 $\pm$   0.16 &  21.67 $\pm$    0.46 & 0.2503 $\pm$  0.0025 & 6   \\
HS 0134+3415      &  7.20 $\pm$   0.19 &  24.31 $\pm$    0.96 & 0.2459 $\pm$  0.0042 & 1   \\
SBS 0917+527      &  7.35 $\pm$   0.29 &  17.69 $\pm$    0.82 & 0.2483 $\pm$  0.0067 & 6   \\
SBS 1152+579      &  7.61 $\pm$   0.18 &  31.43 $\pm$    0.74 & 0.2516 $\pm$  0.0030 & 6   \\
SBS 1533+574A     &  7.65 $\pm$   0.60 &  28.35 $\pm$    2.19 & 0.2422 $\pm$  0.0100 & 6   \\
HS 1028+3843      &  7.81 $\pm$   0.20 &  44.86 $\pm$    1.32 & 0.2449 $\pm$  0.0035 & 1   \\
SBS 0926+606      &  8.27 $\pm$   0.30 &  28.14 $\pm$    1.07 & 0.2466 $\pm$  0.0057 & 6   \\
SBS 1437+370      &  8.43 $\pm$   0.21 &  26.18 $\pm$    0.68 & 0.2517 $\pm$  0.0036 & 6   \\
UM 420            &  8.56 $\pm$   1.00 &  71.44 $\pm$    7.91 & 0.2607 $\pm$  0.0172 & 4   \\
SBS 1222+614      &  9.04 $\pm$   0.26 &  22.26 $\pm$    0.78 & 0.2430 $\pm$  0.0046 & 6   \\
UM 462 SW         &  9.06 $\pm$   0.30 &  27.80 $\pm$    0.99 & 0.2445 $\pm$  0.0049 & 4   \\
HS 0811+4913      &  9.28 $\pm$   0.24 &  32.95 $\pm$    0.92 & 0.2451 $\pm$  0.0040 & 1   \\
SBS 1054+365      &  9.37 $\pm$   0.39 &  30.92 $\pm$    1.50 & 0.2524 $\pm$  0.0067 & 6   \\
UM 448            &  9.89 $\pm$   0.92 &  95.92 $\pm$    8.27 & 0.2513 $\pm$  0.0076 & 4   \\
Mrk 1271          &  9.89 $\pm$   0.32 &  41.17 $\pm$    1.63 & 0.2375 $\pm$  0.0060 & 4   \\
Mrk 59            &  9.90 $\pm$   0.21 &  29.85 $\pm$    0.61 & 0.2416 $\pm$  0.0027 & 6   \\
SBS 0946+558      &  9.93 $\pm$   0.23 &  28.54 $\pm$    0.68 & 0.2516 $\pm$  0.0034 & 6   \\
SBS 0741+535      & 10.23 $\pm$   0.97 &  29.68 $\pm$    2.79 & 0.2465 $\pm$  0.0131 & 6   \\
HS 1214+3801      & 10.44 $\pm$   0.24 &  30.96 $\pm$    0.70 & 0.2435 $\pm$  0.0031 & 1   \\
Mrk 724           & 10.76 $\pm$   0.29 &  29.58 $\pm$    0.77 & 0.2459 $\pm$  0.0036 & 1   \\
HS 0029+1748      & 11.01 $\pm$   0.35 &  45.68 $\pm$    1.40 & 0.2456 $\pm$  0.0042 & 1   \\
NGC 1741          & 11.12 $\pm$   1.42 &  99.73 $\pm$   11.48 & 0.2573 $\pm$  0.0078 & 4   \\
Mrk 67            & 11.16 $\pm$   0.46 &  30.03 $\pm$    1.26 & 0.2494 $\pm$  0.0055 & 1   \\
SBS 1135+581      & 11.26 $\pm$   0.25 &  47.68 $\pm$    0.99 & 0.2439 $\pm$  0.0028 & 6   \\
POX 36            & 11.31 $\pm$   0.56 &  29.54 $\pm$    1.37 & 0.2450 $\pm$  0.0048 & 1   \\
Mrk 5             & 11.39 $\pm$   1.04 &  48.96 $\pm$    4.15 & 0.2487 $\pm$  0.0094 & 4   \\
Mrk 930           & 11.53 $\pm$   0.80 &  46.71 $\pm$    2.96 & 0.2502 $\pm$  0.0079 & 4   \\
SBS 0948+532      & 11.63 $\pm$   0.30 &  51.03 $\pm$    1.30 & 0.2480 $\pm$  0.0037 & 6   \\
UM 439            & 11.67 $\pm$   0.29 &  35.11 $\pm$    1.01 & 0.2444 $\pm$  0.0039 & 1   \\
HS 0924+3821      & 11.78 $\pm$   0.45 &  61.74 $\pm$    2.16 & 0.2482 $\pm$  0.0049 & 1   \\
Mrk 450 (\#2)     & 11.95 $\pm$   0.64 &  48.55 $\pm$    2.43 & 0.2502 $\pm$  0.0070 & 1   \\
SBS 1319+579A     & 12.32 $\pm$   0.35 &  42.17 $\pm$    1.23 & 0.2522 $\pm$  0.0043 & 6   \\
Mrk 750           & 12.90 $\pm$   0.64 &  46.05 $\pm$    2.17 & 0.2436 $\pm$  0.0063 & 4   \\
UM 422            & 12.91 $\pm$   0.36 &  34.09 $\pm$    1.07 & 0.2469 $\pm$  0.0040 & 1   \\
SBS 1533+574B     & 13.13 $\pm$   0.67 &  37.78 $\pm$    1.78 & 0.2466 $\pm$  0.0062 & 6   \\
SBS 1319+579C     & 13.13 $\pm$   1.06 &  52.09 $\pm$    3.76 & 0.2432 $\pm$  0.0069 & 6   \\
Mrk 162           & 13.55 $\pm$   1.10 &  54.68 $\pm$    4.13 & 0.2493 $\pm$  0.0084 & 4   \\
HS 0128+2832      & 13.70 $\pm$   0.33 &  84.10 $\pm$    1.75 & 0.2534 $\pm$  0.0032 & 1   \\
Mrk 1236          & 14.03 $\pm$   0.36 &  65.48 $\pm$    1.48 & 0.2531 $\pm$  0.0033 & 1   \\
Mrk 450 (\#1)     & 14.22 $\pm$   0.39 &  57.95 $\pm$    1.42 & 0.2506 $\pm$  0.0034 & 1   \\
UM 238            & 14.62 $\pm$   0.49 &  79.10 $\pm$    2.43 & 0.2561 $\pm$  0.0047 & 1   \\
HS 0735+3512      & 14.94 $\pm$   0.43 &  52.09 $\pm$    1.39 & 0.2519 $\pm$  0.0041 & 1   \\
Mrk 1063          & 15.24 $\pm$   2.11 &  96.31 $\pm$   11.10 & 0.2490 $\pm$  0.0062 & 1   \\
HS 2359+1659      & 15.24 $\pm$   0.56 &  53.69 $\pm$    1.78 & 0.2503 $\pm$  0.0048 & 1   \\
HS 1311+3628      & 15.25 $\pm$   0.45 &  61.96 $\pm$    1.64 & 0.2514 $\pm$  0.0039 & 1   \\
HS 1213+3636A     & 15.61 $\pm$   0.87 &  52.93 $\pm$    2.64 & 0.2526 $\pm$  0.0042 & 1   \\
UM 396            & 17.08 $\pm$   0.64 &  78.86 $\pm$    2.54 & 0.2528 $\pm$  0.0044 & 1   \\
Mrk 1315          & 17.74 $\pm$   0.41 &  57.57 $\pm$    1.14 & 0.2520 $\pm$  0.0028 & 1   \\
Mrk 1329          & 17.87 $\pm$   0.44 &  59.81 $\pm$    1.23 & 0.2560 $\pm$  0.0028 & 1   \\
HS 0111+2115      & 17.89 $\pm$   2.38 &  65.62 $\pm$    7.52 & 0.2552 $\pm$  0.0098 & 1   \\
Mrk 35            & 19.77 $\pm$   0.58 &  87.53 $\pm$    2.22 & 0.2529 $\pm$  0.0028 & 1   \\
UM 311            & 20.34 $\pm$   2.07 & 111.90 $\pm$    8.62 & 0.2533 $\pm$  0.0063 & 4   \\
  \enddata
\tablerefs{(1) this paper; (2) \citet{I99}; (3) \citet{IT98a}; (4) IT98; (5) 
\citet{ICS01}; (6) ITL97; (7) \citet{G01}; (8) \citet{ICG01}; (9)
\citet{TIF99}; (10) \citet{G03b}; (11) \cite{G03a}. }
  \end{deluxetable}

\clearpage

\begin{deluxetable}{lcrccrc}
 \tabletypesize{\scriptsize}
\tablenum{6}
\tablecolumns{7}
\tablewidth{0pc}
\tablecaption{Maximum Likelihood Linear Regressions}
\tablehead{
\colhead{} &\colhead{Number of}& \multicolumn{2}{c}{Oxygen} && \multicolumn{2}{c}{Nitrogen} \\ \cline{3-4} \cline{6-7}
\colhead{Method}&\colhead{H {\sc ii} Regions}&\colhead{Regression}&\colhead{$\sigma$}&&\colhead{Regression}&\colhead{$\sigma$} }
\startdata
3 He {\sc i} lines\tablenotemark{a,b} &45& 0.2451$\pm$0.0018 + 21$\pm$21(O/H)&0.0048& & 0.2452$\pm$0.0012 + \,\ 603$\pm$372(N/H)&0.0044 \\
3 He {\sc i} lines\tablenotemark{b} &89& 0.2429$\pm$0.0009 + 51$\pm$\,~9(O/H)&0.0040& & 0.2439$\pm$0.0008 +    1063$\pm$183(N/H)&0.0037 \\
5 He {\sc i} lines\tablenotemark{c,d} &7& 0.2421$\pm$0.0021 + 68$\pm$22(O/H)&0.0035& & 0.2446$\pm$0.0016  +    1084$\pm$442(N/H)&0.0040 \\
5 He {\sc i} lines\tablenotemark{c,e} &7& 0.2444$\pm$0.0020 + 61$\pm$21(O/H)&0.0040& & 0.2466$\pm$0.0016 + \,\ 954$\pm$411(N/H)&0.0044 \\
\enddata
\tablenotetext{a}{Data are from IT98.}
\tablenotetext{b}{Only collisional and fluorescent enhancements are taken into
account. We have adopted $T_e$(He {\sc ii}) = $T_e$(O {\sc iii}) and 
$ICF$(He) = 1.}
\tablenotetext{c}{Collisional and fluorescent enhancements of the He {\sc i}
lines, collisional excitation of hydrogen lines,
underlying He {\sc i} stellar absorption and differences between
$T_e$(He {\sc ii}) and $T_e$(O {\sc iii}) are taken into account. $ICF$(He) 
is set to 1.}
\tablenotetext{d}{Calculated with EW$_a$(H8 + He {\sc i} 3889) = 3.0\AA, 
EW$_a$(He {\sc i} 4471) = 0.4\AA, EW$_a$(He {\sc i} 5876) = 0.3 EW$_a$(He {\sc i} 4471),
EW$_a$(He {\sc i} 6678) = EW$_a$(He {\sc i} 7065) = 0.1 EW$_a$(He {\sc i} 4471).}
\tablenotetext{e}{Calculated with EW$_a$(H8 + He {\sc i} 3889) = 3.0\AA, 
EW$_a$(He {\sc i} 4471) = 0.5\AA, EW$_a$(He {\sc i} 5876) = 0.3 EW$_a$(He {\sc i} 4471),
EW$_a$(He {\sc i} 6678) = EW$_a$(He {\sc i} 7065) = 0.1 EW$_a$(He {\sc i} 4471).}
\end{deluxetable}

\clearpage

  \begin{deluxetable}{lccccccc}
  \tabletypesize{\tiny}
  \tablenum{7}
  \tablecolumns{8}
  \tablewidth{0pc}
  \tablecaption{Best model parameters for the restricted sample}
  \tablehead{
  \colhead{Parameter} & \colhead{I Zw 18} &
  \colhead{SBS 0335--052} & \colhead{Mrk 209} &
  \colhead{Mrk 71} & \colhead{NGC 346} &
  \colhead{Mrk 450} & \colhead{UM 311} 
  }
  \startdata
\multicolumn{8}{c}{EW$_a$($\lambda$3889) = 3.0\AA, EW$_a$($\lambda$4471) = 0.4\AA,
EW$_a$($\lambda$5876) = 0.17\AA, EW$_a$($\lambda$6678) = 0.05\AA,
EW$_a$($\lambda$7065) = 0.05\AA} \\ \tableline 
$\chi^2$
& 1.2 & 4.6 & 0.56 
& 3.4$\times$10$^{-3}$ & 4.1 & 0.89 
& 3.4$\times$10$^{-5}$ \\
$\Delta$$I$(H$\alpha$)/$I$(H$\alpha$)
& 0.0495 & 0.0005 & 0.0005 & 0.0375 & 0.0380 & 0.0005 & 0.0210 \\
$T_e$(He {\sc ii})      
& 17150 & 18270 & 15970 & 14130 & 13000 & 11580 &  9439 \\
$T_e$(He {\sc ii})/$T_e$(O {\sc iii})
& 0.902 & 0.902 & 0.990 & 0.902 & 1.000 & 0.992 & 0.970 \\      
$N_e$(He {\sc ii})      
&    10 &   235 &   52 &  193 &   64 &  450 &  79 \\
$\tau$($\lambda$3889)  
&  0.81 & 4.31 & 0.21 & 1.71 & 0.01 & 2.16 & 3.56 \\
$EW_a$/$EW_e$($\lambda$3889)\tablenotemark{a}
& 0.238  &  0.185 &  0.125 &  0.085 &  0.094 &  0.095 &  0.121  \\
$EW_a$/$EW_e$($\lambda$4471)\tablenotemark{a}   
& 0.108  &  0.078 &  0.063 &  0.039 &  0.049 &  0.049 &  0.046 \\
$EW_a$/$EW_e$($\lambda$5876)\tablenotemark{a}
& 0.008  & 0.005  &  0.005 &  0.003 &  0.004 &  0.004 &  0.003  \\
$EW_a$/$EW_e$($\lambda$6678)\tablenotemark{a}
& 0.006  & 0.004  &  0.004 &  0.002 &  0.003 &  0.003 &  0.002  \\
$EW_a$/$EW_e$($\lambda$7065)\tablenotemark{a}
& 0.006  & 0.002  &  0.004 &  0.002 &  0.004 &  0.003 &  0.003  \\
$y^+$($\lambda$3889)
& 0.0743$\pm$0.0060 & 0.0746$\pm$0.0030 & 0.0829$\pm$0.0028 
& 0.0825$\pm$0.0029 & 0.0898$\pm$0.0036 & 0.0831$\pm$0.0042 
& 0.0868$\pm$0.0067 \\
$y^+$($\lambda$4471)
& 0.0829$\pm$0.0061 & 0.0740$\pm$0.0016 & 0.0814$\pm$0.0016
& 0.0826$\pm$0.0014 & 0.0819$\pm$0.0020 & 0.0801$\pm$0.0024 
& 0.0868$\pm$0.0057 \\
$y^+$($\lambda$5876)
& 0.0800$\pm$0.0024 & 0.0780$\pm$0.0012 & 0.0824$\pm$0.0013 
& 0.0826$\pm$0.0013 & 0.0842$\pm$0.0016 & 0.0818$\pm$0.0014 
& 0.0868$\pm$0.0025 \\
$y^+$($\lambda$6678)
& 0.0804$\pm$0.0056 & 0.0774$\pm$0.0016 & 0.0831$\pm$0.0017 
& 0.0826$\pm$0.0015 & 0.0852$\pm$0.0018 & 0.0830$\pm$0.0024 
& 0.0868$\pm$0.0049 \\
$y^+$($\lambda$7065)
& 0.0779$\pm$0.0051 & 0.0762$\pm$0.0014 & 0.0825$\pm$0.0018 
& 0.0826$\pm$0.0015 & 0.0844$\pm$0.0022 & 0.0821$\pm$0.0027 
& 0.0868$\pm$0.0053 \\
$y^+$(mean)
& 0.0795$\pm$0.0018 & 0.0765$\pm$0.0007 & 0.0824$\pm$0.0008 
& 0.0826$\pm$0.0007 & 0.0844$\pm$0.0009 & 0.0818$\pm$0.0010 
& 0.0868$\pm$0.0019 \\
$y^{++}$($\lambda$4686)
& 0.0008$\pm$0.0002 & 0.0024$\pm$0.0001 & 0.0011$\pm$0.0001 
& 0.0008$\pm$0.0001 & 0.0002$\pm$0.0000 & 0.0003$\pm$0.0001 
& \nodata \\
$y$(mean)
& 0.0803$\pm$0.0018 & 0.0789$\pm$0.0007 & 0.0834$\pm$0.0008 
& 0.0834$\pm$0.0007 & 0.0846$\pm$0.0009 & 0.0821$\pm$0.0010 
& 0.0868$\pm$0.0019 \\
$Y$(mean)
& 0.2430$\pm$0.0057 & 0.2399$\pm$0.0022 & 0.2500$\pm$0.0023 
& 0.2499$\pm$0.0021 & 0.2523$\pm$0.0028 & 0.2465$\pm$0.0030 
& 0.2566$\pm$0.0057 \\
12 + log(O/H)\tablenotemark{c}
& 7.28 & 7.39 & 7.78 & 7.97 & 8.01 & 8.16 & 8.36 \\ \tableline
\multicolumn{8}{c}{EW$_a$($\lambda$3889) = 3.0\AA, EW$_a$($\lambda$4471) = 0.5\AA,
EW$_a$($\lambda$5876) = 0.17\AA, EW$_a$($\lambda$6678) = 0.05\AA,
EW$_a$($\lambda$7065) = 0.05\AA} \\ \tableline 
$\chi^2$
& 1.7 & 2.7 & 0.052 
& 0.034 & 3.0 & 0.46 
& 4.5$\times$10$^{-6}$ \\
$\Delta$$I$(H$\alpha$)/$I$(H$\alpha$)
& 0.0495 & 0.0005 & 0.0005 & 0.0475 & 0.0440 & 0.0005 & 0.0390 \\
$T_e$(He {\sc ii})      
& 17150 & 18270 & 16140 & 14110 & 12990 & 11630 &  9112 \\
$T_e$(He {\sc ii})/$T_e$(O {\sc iii})
& 0.902 & 0.902 & 1.000 & 0.902 & 1.000 & 0.996 & 0.938 \\      
$N_e$(He {\sc ii})      
&    10 &   214 &   16 &  157 &   64 &  450 &  151 \\
$\tau$($\lambda$3889)  
&  0.81 & 4.36 & 0.31 & 1.86 & 0.01 & 2.11 & 3.71 \\
$EW_a$/$EW_e$($\lambda$3889)\tablenotemark{a}
& 0.238  &  0.185 &  0.125 &  0.085 &  0.094 &  0.095 &  0.121  \\
$EW_a$/$EW_e$($\lambda$4471)\tablenotemark{a}   
& 0.130  &  0.094 &  0.075 &  0.047 &  0.059 &  0.059 &  0.055 \\
$EW_a$/$EW_e$($\lambda$5876)\tablenotemark{a}
& 0.008  & 0.005  &  0.005 &  0.003 &  0.004 &  0.004 &  0.003  \\
$EW_a$/$EW_e$($\lambda$6678)\tablenotemark{a}
& 0.006  & 0.004  &  0.004 &  0.002 &  0.003 &  0.003 &  0.002  \\
$EW_a$/$EW_e$($\lambda$7065)\tablenotemark{a}
& 0.006  & 0.002  &  0.004 &  0.002 &  0.004 &  0.003 &  0.003  \\
$y^+$($\lambda$3889)
& 0.0742$\pm$0.0060 & 0.0754$\pm$0.0030 & 0.0839$\pm$0.0029 
& 0.0834$\pm$0.0029 & 0.0896$\pm$0.0036 & 0.0827$\pm$0.0041 
& 0.0879$\pm$0.0068 \\
$y^+$($\lambda$4471)
& 0.0848$\pm$0.0063 & 0.0757$\pm$0.0017 & 0.0833$\pm$0.0017
& 0.0838$\pm$0.0015 & 0.0828$\pm$0.0020 & 0.0810$\pm$0.0024 
& 0.0879$\pm$0.0058 \\
$y^+$($\lambda$5876)
& 0.0801$\pm$0.0024 & 0.0787$\pm$0.0013 & 0.0835$\pm$0.0013 
& 0.0839$\pm$0.0013 & 0.0847$\pm$0.0016 & 0.0818$\pm$0.0014 
& 0.0879$\pm$0.0026 \\
$y^+$($\lambda$6678)
& 0.0805$\pm$0.0056 & 0.0777$\pm$0.0016 & 0.0837$\pm$0.0017 
& 0.0837$\pm$0.0015 & 0.0858$\pm$0.0019 & 0.0831$\pm$0.0024 
& 0.0879$\pm$0.0050 \\
$y^+$($\lambda$7065)
& 0.0781$\pm$0.0051 & 0.0768$\pm$0.0014 & 0.0836$\pm$0.0018 
& 0.0837$\pm$0.0015 & 0.0851$\pm$0.0022 & 0.0823$\pm$0.0027 
& 0.0879$\pm$0.0054 \\
$y^+$(mean)
& 0.0798$\pm$0.0018 & 0.0773$\pm$0.0007 & 0.0835$\pm$0.0008 
& 0.0838$\pm$0.0007 & 0.0850$\pm$0.0009 & 0.0820$\pm$0.0010 
& 0.0879$\pm$0.0019 \\
$y^{++}$($\lambda$4686)
& 0.0008$\pm$0.0002 & 0.0024$\pm$0.0001 & 0.0011$\pm$0.0001 
& 0.0008$\pm$0.0001 & 0.0002$\pm$0.0000 & 0.0003$\pm$0.0001 
& \nodata \\
$y$(mean)
& 0.0806$\pm$0.0018 & 0.0797$\pm$0.0007 & 0.0846$\pm$0.0008 
& 0.0846$\pm$0.0007 & 0.0852$\pm$0.0009 & 0.0823$\pm$0.0010 
& 0.0879$\pm$0.0019 \\
$Y$(mean)
& 0.2437$\pm$0.0057 & 0.2417$\pm$0.0022 & 0.2525$\pm$0.0024 
& 0.2525$\pm$0.0022 & 0.2536$\pm$0.0028 & 0.2469$\pm$0.0030 
& 0.2590$\pm$0.0058 \\
12 + log(O/H)\tablenotemark{c}
& 7.28 & 7.39 & 7.77 & 7.98 & 8.01 & 8.16 & 8.41 
  \enddata
\tablenotetext{a}{Ratio of the absorption-to-emission line equivalent widths.}
\tablenotetext{b}{Correction of $Y$ for systematic effects.}
\tablenotetext{c}{Calculated with $T_e$(He {\sc ii}).}
  \end{deluxetable}

\clearpage

\begin{figure*}
\figurenum{1}
\epsscale{0.9}
\plotone{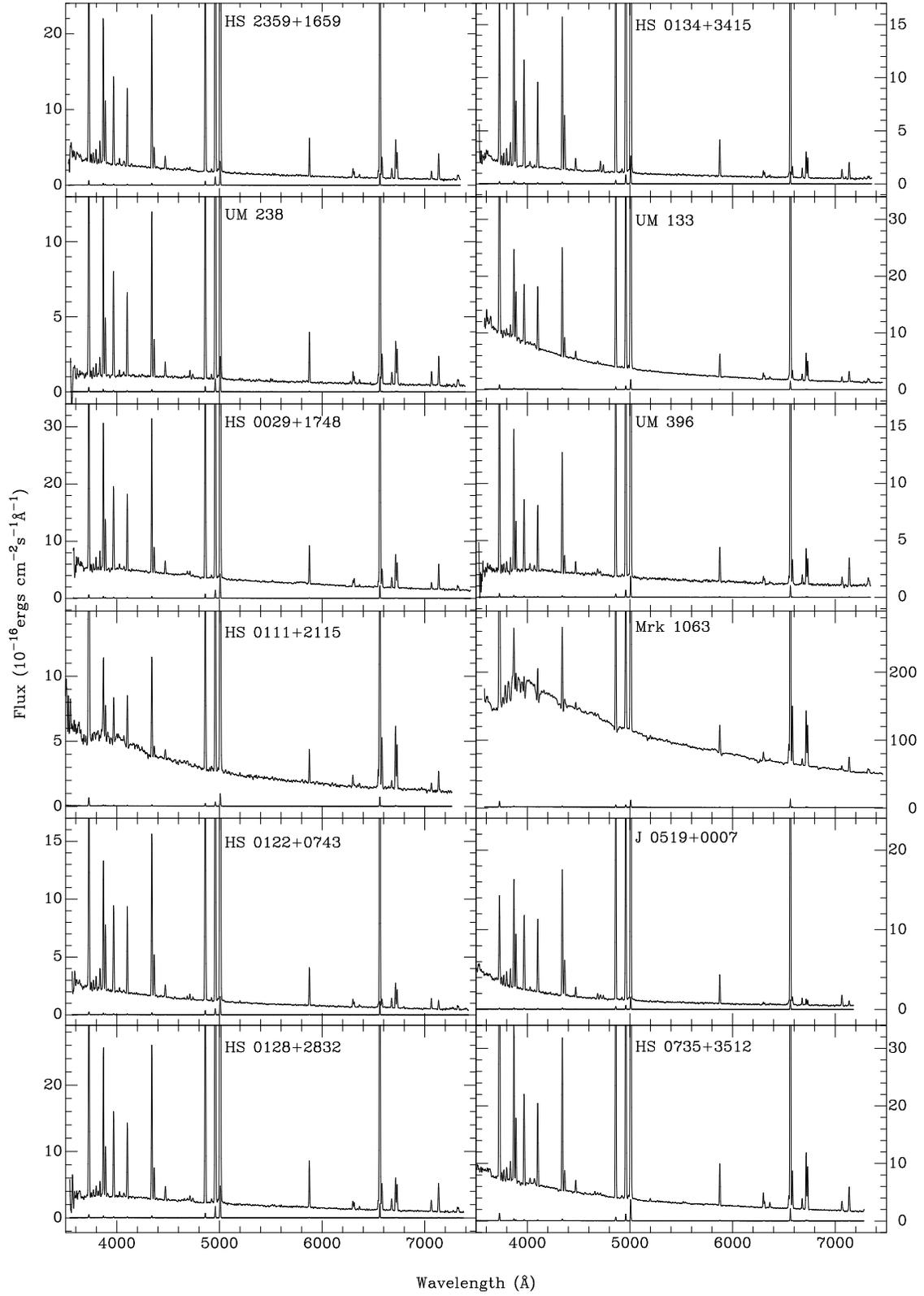}
\figcaption{Mayall 4-m telescope spectra of 33 H {\sc ii} regions in 31 
blue compact galaxies.\label{fig1}}
\end{figure*}

\clearpage

\begin{figure*}
\figurenum{1}
\epsscale{0.9}
\plotone{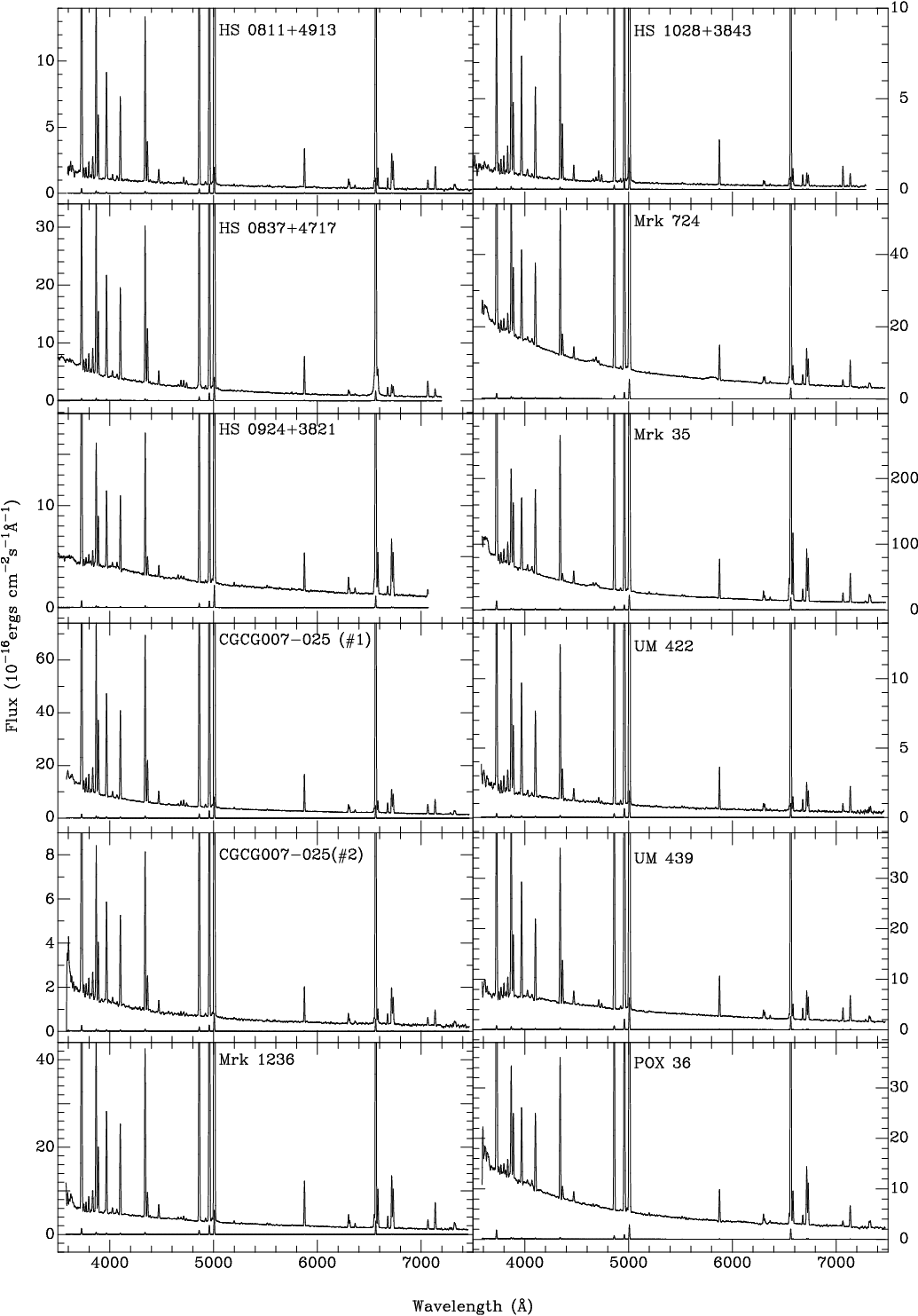}
\figcaption{Continued.}
\end{figure*}

\clearpage

\begin{figure*}
\figurenum{1}
\epsscale{0.9}
\plotone{f1c.ps}
\figcaption{Continued.}
\end{figure*}

\clearpage

\begin{figure*}
\figurenum{2}
\epsscale{1.1}
\plottwo{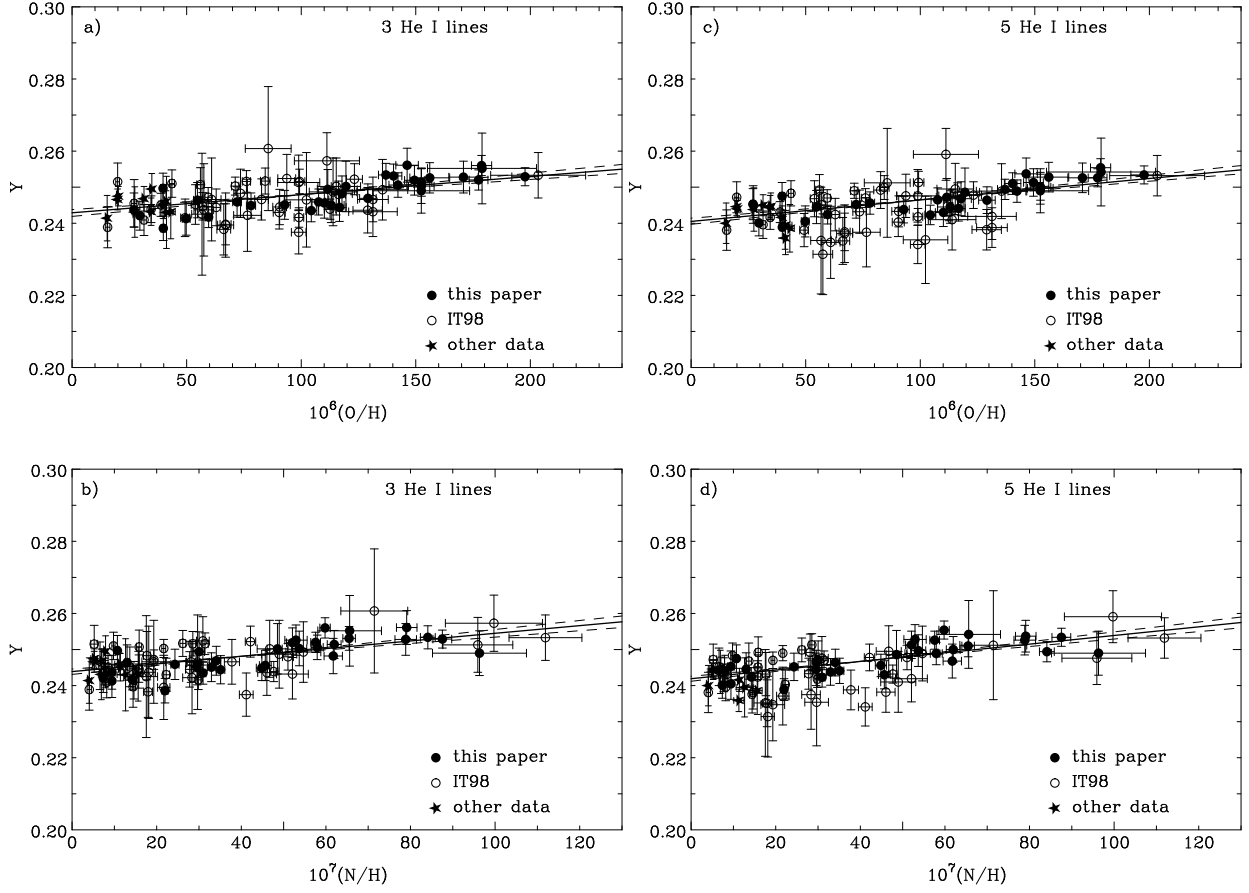}{f2b.ps}
\figcaption{Linear regressions of the helium mass fraction $Y$ vs. oxygen and
nitrogen abundances for a total of 82 H {\sc ii} regions in 76 blue compact 
galaxies. In panels a) and b), $Y$ was derived using the 3 $\lambda$4471, 
$\lambda$5876 and $\lambda$6678 He {\sc i} lines, and in panels c) and d),
$Y$ was derived using the 5 $\lambda$3889, $\lambda$4471, 
$\lambda$5876, $\lambda$6678 and $\lambda$7065 He {\sc i} lines.\label{fig2}}
\end{figure*}

\clearpage

\begin{figure*}
\figurenum{3}
\epsscale{1.1}
\plottwo{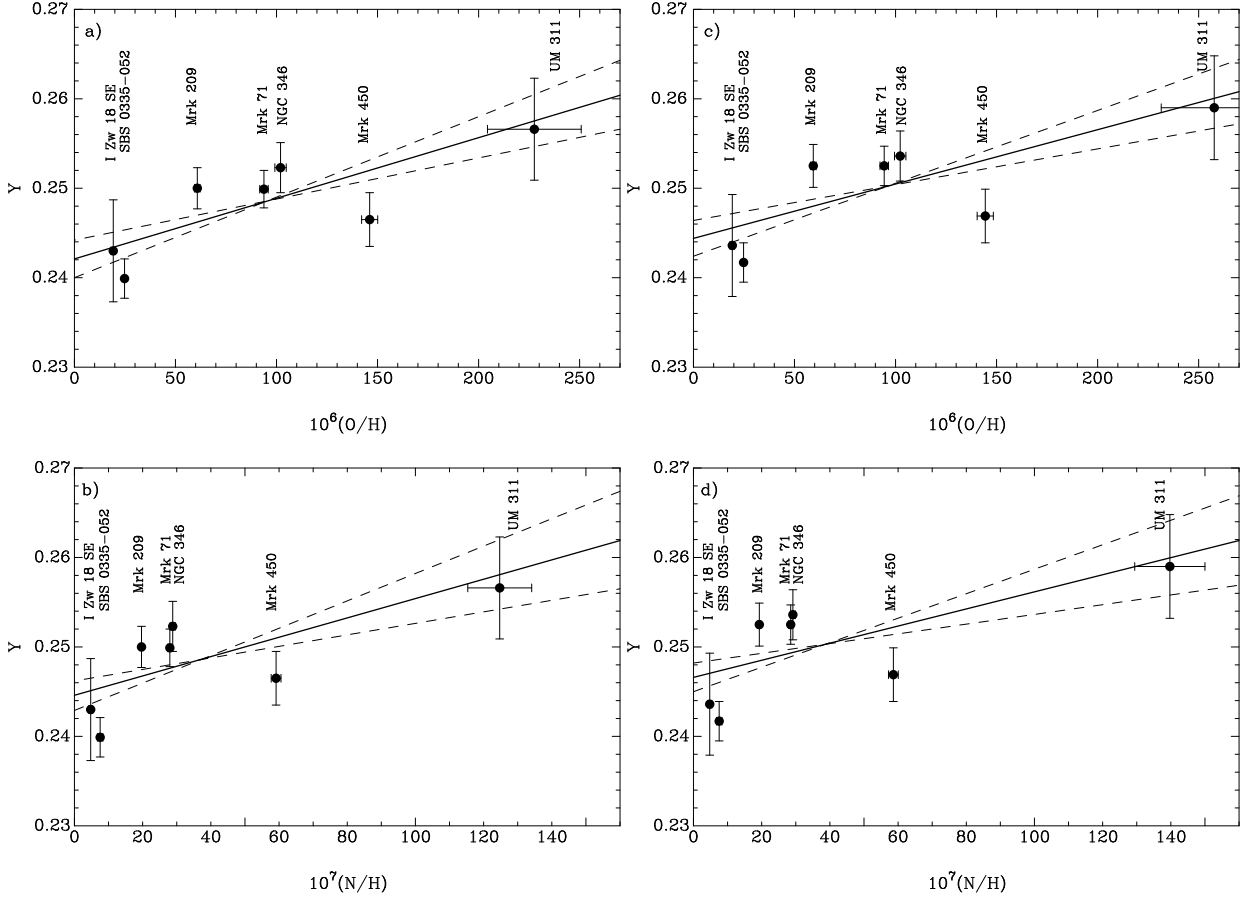}{f3b.ps}
\figcaption{$Y$ -- O/H (a,c) and $Y$ -- N/H (b,d) linear regressions 
(solid lines)
for seven H {\sc ii} regions. Their helium mass fraction $Y$ has been corrected
for known systematic effects by a $\chi^2$ minimization procedure (see \S6).
The filled circles represent the solution with the lowest $\chi^2$.
Oxygen and nitrogen abundances for all points are calculated by 
setting the electron temperatures in the O {\sc iii} zone equal to
$T_e$(He {\sc ii}), the latter being derived from $\chi^2$ minimization.
The equivalent widths of the absorption lines adopted in (a) and (b)
are EW$_a$($\lambda$3889) = 3.0\AA, EW$_a$($\lambda$4471) = 0.4\AA, EW$_a$($\lambda$5876) 
= 0.17\AA, EW$_a$($\lambda$6678) = EW$_a$($\lambda$7065) = 0.05\AA. The 
corresponding
values in (c) and (d) are 
EW$_a$($\lambda$3889) = 3.0\AA, EW$_a$($\lambda$4471) = 0.5\AA, EW$_a$($\lambda$5876) 
= 0.17\AA, EW$_a$($\lambda$6678) = EW$_a$($\lambda$7065) = 0.05\AA.
\label{fig3}}
\end{figure*}


\begin{table}
\dummytable\label{tab1}
\end{table}

\begin{table}
\dummytable\label{tab2}
\end{table}

\begin{table}
\dummytable\label{tab3}
\end{table}

\begin{table}
\dummytable\label{tab4}
\end{table}

\begin{table}
\dummytable\label{tab5}
\end{table}

\begin{table}
\dummytable\label{tab6}
\end{table}

\begin{table}
\dummytable\label{tab7}
\end{table}

\end{document}